\documentclass[english,pra,lengthcheck,final,superscriptaddress,longbibliography]{revtex4-2}
\usepackage[T1]{fontenc}
\usepackage[latin9]{inputenc}
\usepackage{color}
\usepackage{babel}
\usepackage{amsmath}
\usepackage{amssymb}
\usepackage{graphicx}
\usepackage[unicode=true,pdfusetitle,
 bookmarks=true,bookmarksnumbered=false,bookmarksopen=false,
 breaklinks=false,pdfborder={0 0 1},backref=false,colorlinks=true]
 {hyperref}
\begin{document}

\title{Decoherence scaling transition in the dynamics of quantum information
scrambling}

\author{Federico D. Domínguez}

\affiliation{Centro Atómico Bariloche, CONICET, CNEA, S. C. de Bariloche, Argentina.}

\author{María Cristina Rodríguez}

\affiliation{Centro Atómico Bariloche, CONICET, CNEA, S. C. de Bariloche, Argentina.}

\affiliation{Instituto Balseiro, CNEA, Universidad Nacional de Cuyo, S. C. de
Bariloche, Argentina.}

\author{Robin Kaiser}

\affiliation{Université Côte d\textquoteright Azur, CNRS, INPHYNI, F-06560, Valbonne,
France.}

\author{Dieter Suter}

\affiliation{Fakultät Physik, Technische Universität Dortmund, D-44221, Dortmund,
Germany.}

\author{Gonzalo A. Álvarez}
\email{Corresponding author: gonzalo.alvarez@cab.cnea.gov.ar}

\affiliation{Centro Atómico Bariloche, CONICET, CNEA, S. C. de Bariloche, Argentina.}

\affiliation{Instituto Balseiro, CNEA, Universidad Nacional de Cuyo, S. C. de
Bariloche, Argentina.}

\affiliation{Instituto de Nanociencia y Nanotecnologia, CNEA, CONICET, S. C. de
Bariloche, 8400, Argentina}
\begin{abstract}
Reliable processing of quantum information for developing quantum
technologies requires precise control of out-of-equilibrium many-body
systems. This is a highly challenging task as the fragility of quantum
states to external perturbations increases with the system-size. Here,
we report on a series of experimental quantum simulations that quantify
the sensitivity of a controlled Hamiltonian evolution to perturbations
that drive the system away from the targeted evolution. Based on out-of-time
ordered correlations, we demonstrate that the decay-rate of the process
fidelity increases with the effective number $K$ of correlated qubits
as $K^{\alpha}$. As a function of the perturbation strength, we observe
a decoherence scaling transition of the exponent $\alpha$ between
two distinct dynamical regimes. In the limiting case below the critical
perturbation strength, the exponent $\alpha$ drops sharply below
1, and there is no inherent limit to the number of qubits that can
be controlled. This resilient quantum feature of the controlled dynamics
of quantum information is promising for reliable control of large
quantum systems.
\end{abstract}
\maketitle

\section{Introduction}

The characterization and understanding of the complex dynamics of
interacting many-body quantum systems is an outstanding problem in
physics \citep{Eisert2015,Abanin2019}. They play a crucial role in
condensed matter physics, cosmology, quantum information processing
and nuclear physics \citep{Martinez2016,Swingle2018,Friis2018,lewis-swan2019dynamics}.
A particularly urgent issue is the reliable control of many-body quantum
systems, as it is perhaps the most important step towards the development
and deployment of quantum technologies \citep{Eisert2015,Zhang2017,Bernien2017,Neill2018}.
Their control is never perfect and the fragility of quantum states
to perturbations increases with the system size \citep{krojanski2004scaling,alvarez2015localization,suter2016colloquium}.
Accordingly, information processing with large quantum systems remains
a challenging task. It is therefore of paramount importance to reduce
the sensitivity to perturbations, particularly for large systems,
to minimize the loss of quantum information. As we show here, achieving
this goal may be more realistic than it is currently assumed:  we
demonstrate that the sensitivity of a quantum evolution to imperfections
in the control operation  can become qualitatively smaller, provided
that perturbation strengths are below a certain threshold.

Perturbations to the control Hamiltonian due to uncontrolled degrees
of freedom, degrade the quantum information in a process generally
known as decoherence. Mitigating this effect has been the goal of
numerous studies to allow information storage by protecting quantum
states from perturbations \citep{suter2016colloquium,Wang2017}.
However, characterizing and controlling decoherence effects during
the dynamics of quantum information remain challenging tasks, since
out-of-equilibrium many-body physics is involved \citep{Trotzky2012,Schindler2013,Zhang2017,lewis-swan2019dynamics,Landsman2019}.
Theoretical and experimental approaches were developed to reduce decoherence
in few-body systems \citep{Sar2012,Souza2012,Taminiau2014,Zhang2015,suter2016colloquium}.
Extending these approaches to larger quantum systems is not a straightforward
scaling operation, since the evolution in these systems generates
high-order quantum correlations that are spread over degrees of freedom
of many qubits. Controlling and probing these correlations was tackled
only recently \citep{alvarez2015localization,Schweigler2017,Friis2018,Lukin2019,Landsman2019,Brydges2019}.
Novel techniques are therefore required to address this task, in particular
with quantum simulations \citep{buluta2009quantum,Schindler2013,georgescu2014quantum,alvarez2015localization,Zhang2017,Bernien2017}.

The dynamics of the build-up of many-body quantum superpositions was
initially measured within nuclear magnetic resonance (NMR) by observing
multiple quantum coherences (MQC) \citep{baum1985multiple}. MQCs
are relatively easy to characterize as they do not require a full
quantum state tomography and the coherence order provides a hard lower
bound on the number of correlated particles (spins) \citep{alvarez2015localization,garttner2017measuring,garttner2018relating,lewis-swan2019dynamics}.
MQCs can be useful tools to measure the sensitivity of controlled
dynamics to perturbations \citep{alvarez2010nmr,alvarez2015localization}
combined with time-reversal of quantum evolutions that leads to a
Loschmidt Echo \citep{peres1984stability,pastawski2000nuclear,jacquod2009decoherence}.
Loschmidt echoes and MQC evidence out-of-time order correlations (OTOC)
\citep{Swingle2018,lewis-swan2019dynamics}, as they measure the scrambling
of the information over a large system from an initially localized
state \citep{garttner2017measuring,garttner2018relating,yan2019information}.
They are therefore promising tools for finding answers to open questions
related to quantum chaos \citep{maldacena2016bound,li2017measuring,niknam2018sensitivity},
irreversibility \citep{yan2019information,sanchez2020perturbation},
thermalization \citep{wei2019emergent} and entanglement \citep{garttner2018relating}.
Hence, these OTOCs trigger a broad interest in diverse fields of physics,
such as condensed matter and quantum gravity \citep{garttner2018relating,niknam2018sensitivity,li2017measuring,wei2018exploring,garttner2017measuring,fan2017out,garcia2018chaos,wei2019emergent},
opening new avenues for understanding the dynamics of quantum information
in complex systems \citep{Swingle2018,lewis-swan2019dynamics}.

Here, we use tools of solid-state NMR to assess the sensitivity to
perturbations of a controlled quantum dynamics in a many-body system.
We drive the system away from equilibrium by suddenly imposing on
it an experimentally controllable Hamiltonian that does not commute
with the initial condition and that can be inverted, in order to drive
the system forward or backward in time. The forward motion causes
the quantum information to spread over a large system (with thousands
of particles), but in the case where the inversion of the Hamiltonian
is perfect, the system returns exactly to the initial state \textendash this
is known as a Loschmidt Echo \citep{peres1984stability,pastawski2000nuclear,jacquod2009decoherence}.

In practice, the inversion of the Hamiltonian is never perfect, and
the deviations result in imperfect return to the initial condition
and therefore to a reduction of the echo signal, which is proportional
to the overlap between the initial and final state. Here, we study
the effect of such deviations from the ideal Hamiltonian by adding
perturbations with variable strength $p$ and measuring their effect
on the evolution. This sets a paradigmatic model system where initial
information stored on local states spreads over a spin-network of
about 5000 spins. This information spreading process is called scrambling
\citep{maldacena2016bound,Swingle2018,lewis-swan2019dynamics}, to
indicate that the local initial condition can no longer be accessed
by local measurements. We experimentally design an OTOC measure to
probe high order quantum correlations and compare the scrambling of
information from the initial state by the ideal and the perturbed
quantum dynamics. This is done by implementing a Loschmidt Echo with
a forward evolution driven by the perturbed Hamiltonian and a backward
evolution driven by the ideal one, so as to quantify the difference
between the scrambling dynamics. We derive an OTOC that defines an
effective cluster size, the number of correlated spins $K$ over which
the information was spread by the ideal control Hamiltonian. We demonstrate
that the fidelity decay rate of the controlled dynamics \textendash measured
with the Loschmidt Echo\textendash{} increases with the instantaneous
cluster-size $K$, as a power law $\propto K^{\alpha}$, with $\alpha$
depending on the perturbation strength $p$. Strikingly, our results
evidence two qualitatively different fidelity decay regimes with distinctive
scaling laws associated with a sudden change of the exponent $\alpha$.
For perturbations larger than a given threshold, the controlled dynamics
is localized, as manifested by a saturation of the cluster-size growth
$K(t)$. This imposes a limit on the number of qubits that can be
controlled during a quantum operation. However, for perturbations
lower than the threshold, the cluster-size $K$ grows indefinitely
and the exponent $\alpha$ drops abruptly, making the quantum dynamics
of large systems qualitatively more resilient to perturbations. This
sudden sensitivity reduction to perturbations is a promising quantum
feature that may be used to implement reliable quantum information
processing with many-body systems for novel quantum technologies and
for studying quantum information scrambling.

\section{Quantum information dynamics}

We perform experimentally all quantum simulations on a Bruker Avance
III HD 9.4T WB NMR spectrometer with a $^{\text{1}}$H resonance frequency
of $\omega_{z}=400.15$ MHz. We consider the spins of the Hydrogen
nuclei of polycrystalline adamantane, where the strength of the average
dipolar interaction can be determined from the full-width-half-maximum
of the resonance line 13 kHz. They constitute an interacting many-body
system of equivalent spins $I=1/2$ in a strong magnetic field. In
the rotating frame of reference, the Hamiltonian reduces to \citep{slichter2013principles}
\begin{equation}
\mathcal{H}_{dd}=\sum_{i<j}d_{ij}\left[2I_{z}^{i}I_{z}^{j}-(I_{x}^{i}I_{x}^{j}+I_{y}^{i}I_{y}^{j})\right],\label{eq:dipolar_interaction}
\end{equation}
where $I_{x}^{i},I_{y}^{i}\mbox{ and }I_{z}^{i}$ are the spin operators
and $d_{ij}$ the spin-spin coupling strengths that scale with the
distance between spins $\propto1/r_{ij}^{3}$. The dipolar interaction
$\mathcal{H}_{dd}$ is truncated to the part that commutes with the
stronger Zeeman interaction ($\omega_{z}\gg d_{ij}$), as the effects
of the non-commuting part are negligible.

The NMR quantum simulations start from the high-temperature thermal
equilibrium state $\rho(0)\approx\left(\mathbb{I}+\frac{\hbar\omega_{z}}{k_{\mathrm{B}}T}I_{z}\right)/\mathrm{Tr}\left\{ \mathbb{I}\right\} $,
where $I_{z}=\sum_{i}I_{z}^{i}$ commutes with the Hamiltonian $\mathcal{H}_{dd}$
\citep{slichter2013principles}. The unity operator $\mathbb{I}$
does not contribute to an observable signal (see Appendix \ref{sec:Initial-state.}).
In this state, the spins are uncorrelated and form the ensemble of
local states that we consider as the initial local information.

To spread the local information, we drive the system out of equilibrium
with the evolution operator $U_{0}(t)=e^{-it\mathcal{H}_{0}}$, with
the double-quantum Hamiltonian

\begin{equation}
\mathcal{\mathcal{H}}_{0}=-\sum_{i<j}d_{ij}\left[I_{x}^{i}I_{x}^{j}-I_{y}^{i}I_{y}^{j}\right]\label{flip-flip}
\end{equation}
as the ideal \textendash non-perturbed\textendash{} Hamiltonian. This
Hamiltonian flips simultaneously two spins with the same orientation.
Accordingly, the $z$-component of the magnetization $M_{z}$ changes
by $M=\Delta M_{z}=\pm2.$ At the same time, the number of correlated
spins $K$ changes by $\Delta K=\pm1$ \citep{Munowitz1987} (see
Appendix \ref{sec:Determination-of-the}). The coherence order $M=M_{z,j}-M_{z,i}$,
classifies the coherences $|M_{z,i}\rangle\langle M_{z,j}|$ of the
density matrix, where $I_{z}\left|M_{z,i}\right\rangle =M_{z,i}\left|M_{z,i}\right\rangle $.
The change of coherence order allows to probe high-order spin correlations
associated with the number of correlated spins that witness the information
spreading over the system from the initial ensemble of localized states
\citep{baum1985multiple,Munowitz1987} (see Appendix \ref{sec:Determination-of-the}).

To quantify the sensitivity to perturbations of the controlled quantum
dynamics, we control the deviation from $\mathcal{H}_{0}$ with the
dimensionless perturbation strength $p$ of the Hamiltonian
\begin{equation}
\mathcal{H}(p)=(1-p)\mathcal{\mathcal{H}}_{0}+p\Sigma.\label{eq:hpert-1-1-1}
\end{equation}
Here $\Sigma$ is a perturbation Hamiltonian. The Hamiltonian $\mathcal{H}$
is engineered with average Hamiltonian techniques using a NMR pulse
sequence \citep{alvarez2010nmr,alvarez2015localization} (see Appendix
\ref{sec:Hamiltonian-engineering.}). We consider the effect of two
different perturbations: i) a two spin operator perturbation given
by the dipolar Hamiltonian $\Sigma=\mathcal{H}_{dd}$ and ii) a single
spin operator perturbation given by a longitudinal offset field $\Sigma=\mathcal{H}_{z}=\Delta\omega_{z}I_{z}$
\citep{wei2019emergent}. Both perturbations induce a controlled
relative dephasing with respect to the ideal evolution that produces
decoherence effects. 

\section{Fidelity and Loschmidt Echo}

The observable in the experiments is the magnetization operator $I_{z}$.
Since the trace of this observable is zero, the unity term $\mathbb{I}$
in the initial state $\rho(0)$ does not contribute to its expectation
value, and our observable signal $\mathrm{Tr}\left[I_{z}\rho(t)\right]\ensuremath{\propto\mathrm{Tr}}\left[I_{z}I_{z}(t)\right]$
gives the $I_{z}$ evolution. Therefore, we can quantify the deviation
between the actual driven state $I_{z}(t)=U_{p}(t)I_{z}U_{p}^{\dagger}(t)$
and the ideally driven state $I_{z}^{0}(t)=U_{0}(t)I_{z}U_{0}^{\dagger}(t)$,
where $U_{p}(t)=e^{-it\mathcal{H}(p)}$ is the perturbed operation
and $U_{0}(t)=e^{-it\mathcal{H}_{0}}$ the ideal control operation.
The instantaneous state fidelity is defined by the inner-product between
$I_{z}(t)$ and $I_{z}^{0}(t)$ that is determined after a proper
normalization of the NMR signal
\begin{equation}
f(t)=\mathrm{Tr}\left[I_{z}(t)I_{z}^{0}(t)\right]/\mathrm{Tr}\left(I_{z}^{2}\right),\label{eq:fidelity}
\end{equation}
where the factor $\mathrm{Tr}(I_{z}^{2})^{-1}$ ensures that $f(0)=1$
(see Appendix \ref{sec:Fidelity.}).  
\begin{figure*}
\begin{centering}
\includegraphics[width=1\textwidth]{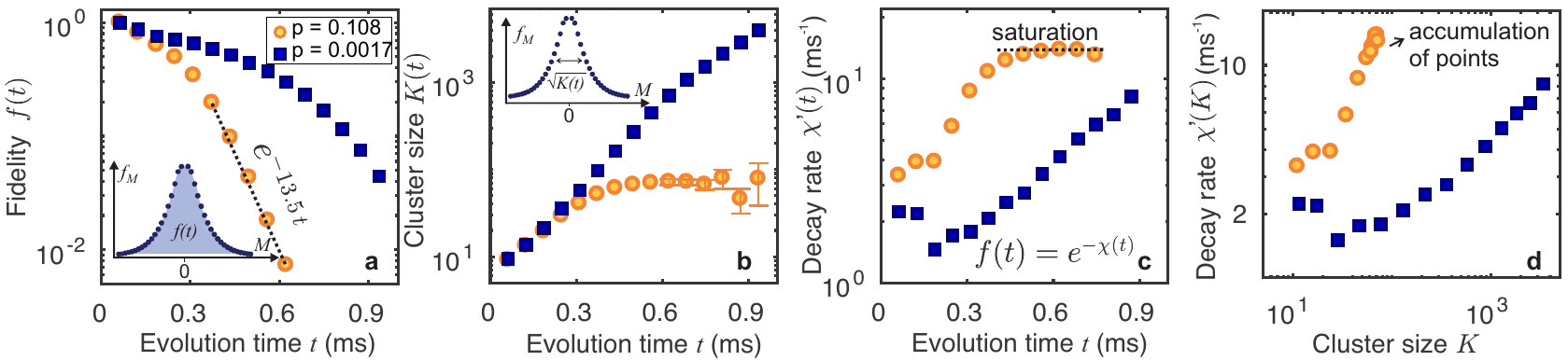}
\par\end{centering}
\caption{Time evolution of the controlled-dynamics' fidelity and the corresponding
effective cluster-size of correlated spins as a measure of the scrambling
of information. (a) The fidelity decay $f(t)=\text{Tr}\left[I_{z}(t)I_{z}^{0}(t)\right]/\mathrm{Tr}\left(I_{z}^{2}\right)$
is shown for two perturbation strengths, for the perturbation Hamiltonian
$\Sigma=\mathcal{H}_{dd}$. The strongest perturbation shows an exponential
decay law for times > 0.3 ms. (inset) A MQC-fidelity $f_{M}(t)$ between
the perturbed dynamics $I_{z}(t)$ and the ideal \textendash non-perturbed\textendash{}
dynamics $I_{z}^{0}(t)$ as a function of the coherence order $M$.
The enclosed area gives the global fidelity $f(t)$. (b) Evolution
of the cluster-size of correlated spins $K(t)$ determined from the
second moment of the MQC-fidelity (inset). The number of correlated
spins $K(t)$ defines the ``coherence length'' on which the density
matrices are comparable. For the weakest perturbations, the cluster
size grows indefinitely, while for the strongest ones, $K(t)$ reaches
a stationary value \textendash an effect we call localization\textendash .
(c) The instantaneous decoherence rate $\chi'(p,t)=\frac{d\chi}{dt}(p,t)$
of the fidelity $f(t)$ as a function of time $t$. The exponential
decay regimen of $f(t)$ is manifested here when the decoherence rate
$\chi'(t)$ achieves a constant value. (d) The instantaneous decoherence
rate $\chi'$ as a function of the cluster-size $K$. The plateau
of $\chi'(t)$ that appears when the cluster size $K(t)$ localizes
in (c), here is manifested by the accumulation of points at the end
of the curve $\chi'(K)$. \label{fig: f_k_and_chi}}
\end{figure*}

This fidelity is identical to the Loschmidt Echo \citep{peres1984stability,pastawski2000nuclear,jacquod2009decoherence}
by choosing the observable magnetization $I_{z}$ equal to the initial
magnetization. We first evolve the system with the perturbed evolution
operator $U_{p}(t)$ and then we time-reverse the evolution with the
unperturbed evolution operator $U_{0}^{\dagger}(t)$. The observable
signal $\propto\mathrm{Tr}\left(U_{0}^{\dagger}U_{p}I_{z}U_{p}^{\dagger}U_{0}\cdot I_{z}\right)$
gives the many-body Loschmidt-Echo 
\begin{equation}
f(t)=\mathrm{Tr}\left(U_{0}^{\dagger}U_{p}I_{z}U_{p}^{\dagger}U_{0}\cdot I_{z}\right)/\mathrm{Tr}\left(I_{z}^{2}\right)
\end{equation}
that is equal to the fidelity after considering cyclic permutations
(see Appendix \ref{sec:Fidelity.}).

\section{Multiple-Quantum Fidelity and OTOC}

We perform a partial tomography of the density matrix fidelity by
applying a rotation operation $\phi_{z}=e^{i\ensuremath{\phi I_{z}}}$
between the forward $U_{p}(t)$ and backward evolution $U_{0}^{\dagger}(t)$.
The global fidelity becomes
\begin{multline}
f_{\phi}(t)=\mathrm{Tr}\left(U_{0}^{\dagger}\phi_{z}U_{p}I_{z}U_{p}^{\dagger}\phi_{z}^{\dagger}U_{0}\cdot I_{z}\right)/\mathrm{Tr}\left(I_{z}^{2}\right)\\
=\mathrm{Tr}\left[I_{z}(t)\phi_{z}^{\dagger}I_{z}^{0}(t)\phi_{z}\right]/\mathrm{Tr}\left(I_{z}^{2}\right)\\
={\textstyle \sum_{M}}e^{i\phi M}f_{M}(t)/\mathrm{Tr}\left(I_{z}^{2}\right),\label{eq:OTOC}
\end{multline}
where we decompose it into the partial MQC inner-products 
\begin{equation}
f_{M}(t)=\mathrm{Tr}\left[I_{z,M}(t)I_{z,M}^{0}(t)\right]/\mathrm{Tr}\left(I_{z}^{2}\right)
\end{equation}
of different coherence orders $M$. The overlap $f_{M}(t)$ quantifies
the deviation of the density operator elements with a given $M$ of
the perturbed evolution from the ideal ones (see Appendix \ref{sec:Determination-of-the}).

If the perturbation strength $p=0$, Equation (\ref{eq:OTOC}) gives
a conventional OTOC
\begin{align}
f_{\phi}(p=0,t) & =\left\langle I_{z}^{0}(t)\phi_{z}^{\dagger}I_{z}^{0}(t)\phi_{z}\right\rangle _{\beta=0}\\
 & =\left\langle \phi_{z}(t)I_{z}\phi_{z}^{\dagger}(t)I_{z}\right\rangle _{\beta=0},
\end{align}
with $\phi_{z}(t)=U_{0}(t)\phi_{z}U_{0}^{\dagger}(t)$. Here $\left\langle \cdot\right\rangle _{\beta=0}=\mathrm{Tr\left(\cdot\right)/\mathrm{Tr}}\left(I_{z}^{2}\right)$
is an expectation value normalized to its value at $t=0$ if the system
is assumed at infinite temperature (see Appendix \ref{sec:OTOCs-and-the}).
It quantifies the scrambling into the system of the local information
stored in the initial state $\rho(0)$  \citep{garttner2017measuring,lewis-swan2019dynamics}.
The components $f_{M}(p=0,t)=\mathrm{Tr}\left[I_{z,M}^{0}(t)I_{z,M}^{0}(t)\right]/\mathrm{Tr}\left(I_{z}^{2}\right)$
are the amplitudes of the MQC spectrum representing the distribution
of coherences (non-diagonal terms in the eigenbasis of $I_{z}$) of
the density matrix that were built by the control Hamiltonian $\mathcal{H}_{0}$
\citep{baum1985multiple,alvarez2015localization}. The second-moment
of the MQC spectrum $f_{M}(p=0,t)$ provides the average cluster-size
of correlated spins 
\begin{align}
\tfrac{K_{0}(t)}{2} & =\sum_{M}M^{2}f_{M}(p=0,t)\\
 & =\mathrm{\mathrm{Tr}}\left([I_{z}^{0}(t),I_{z}]^{\dagger}[I_{z}^{0}(t),I_{z}]\right)/\mathrm{Tr}\left(I_{z}^{2}\right)
\end{align}
at the evolution time $t$ \citep{baum1985multiple,alvarez2015localization,garttner2017measuring,garttner2018relating}
(see Appendix \ref{sec:OTOCs-and-the}). The expression $\mathrm{\mathrm{Tr}}\left([I_{z}^{0}(t),I_{z}]^{\dagger}[I_{z}^{0}(t),I_{z}]\right)/\mathrm{Tr}\left(I_{z}^{2}\right)=\left\langle [I_{z}^{0}(t),I_{z}]^{\dagger}[I_{z}^{0}(t),I_{z}]\right\rangle _{\beta=0}$
is a commutator OTOC that quantifies the degree by which the initially
commuting operators $I_{z}^{0}(t)$ and $I_{z}$ fail to commute at
time $t$ due to the scrambling of information induced by the spin-spin
interactions of $\mathcal{H}_{0}$ \citep{garttner2017measuring,garttner2018relating}\emph{.}

Considering the perturbed evolution ($p\ne0$), the fidelity $f_{\phi}(t)$
is a more general OTOC that quantifies the deviation of the information
scrambling induced by $\mathcal{H}(p)$ with respect to the one driven
by $\mathcal{H}_{0}$. This is seen from the second moment of $f_{M}(t)$,
\begin{equation}
\sum_{M}M^{2}f_{M}(t)=\mathrm{Tr}\left([I_{z}(t),I_{z}]^{\dagger}[I_{z}^{0}(t),I_{z}]\right)/\mathrm{Tr}\left(I_{z}^{2}\right),\label{eq:GOTOC-1}
\end{equation}
that based on the inner-product between the commutators $[I_{z}(t),I_{z}]$
and $[I_{z}^{0}(t),I_{z}]$ gives the degree of non-commutation
shared by the evolved states $I_{z}^{0}(t)$ and $I_{z}(t)$ with
respect to $I_{z}$ (see Appendix \ref{sec:OTOCs-and-the}). Since
$\sum_{M}f_{M}=f(t)=\mathrm{Tr}\left[I_{z}(t)I_{z}^{0}(t)\right]/\mathrm{Tr}\left(I_{z}^{2}\right)$
decays as a function of time, the cluster size of correlated spins
is determined from the normalized second moment
\begin{equation}
K(t)=\tfrac{2\sum_{M}M^{2}f_{M}(t)}{\sum_{M}f_{M}(t)}.
\end{equation}
As the perturbation Hamiltonians $\mathcal{H}_{dd}$ and $\mathcal{H}_{z}$
do not generate MQC by themselves, the OTOC of Eq. (\ref{eq:GOTOC-1})
provides the scrambling of information by the spin-spin interactions
of $\mathcal{H}_{0}$ that survived the perturbation effects. Based
on the second moment of $f_{M}(t)$, $K(t)$ defines a ``coherence
length'' between the two scrambling dynamics of information in terms
of an \emph{average hamming weight} \citep{wei2018exploring,niknam2018sensitivity,wei2019emergent}
for the fidelity of the density matrix. Therefore $K(t)$ quantifies
how comparable the perturbed and unperturbed density matrix dynamics
are as a function of the coherence order $M$. This coherence length
$K(t)$ defines the effective cluster-size of correlated spins on
which the density matrices are comparable based on the inner-product
$f_{M}(t)=\mathrm{Tr}\left[I_{z,M}(t)I_{z,M}^{0}(t)\right]/\mathrm{Tr}\left(I_{z}^{2}\right)$
as a kind of fidelity in Eq. (\ref{eq:GOTOC-1}). In experimental
implementations of quantum simulations, there are always uncontrolled
perturbations \citep{suter2016colloquium} (see Appendix \ref{sec:Intrinsic-Decoherence-effects.}).
These interactions add extra terms to $\Sigma$ that are responsible
for the fidelity decay even when the controlled perturbation is $p=0$.
This can be interpreted as the effective perturbation strength is
no null.

We measure the time evolution of the MQC-fidelities $f_{M}(t)$ for
different perturbations to determine the global fidelity $f(t)$ and
the effective cluster-size $K(t)$. Both are shown in Fig. \ref{fig: f_k_and_chi}(a)
and (b), respectively, for a weak ($p=0.0017$) and a strong perturbation
strength ($p=0.108$) when $\Sigma=\mathcal{H}_{dd}$. The fidelity
decays faster as a function of time with increasing perturbation strength.
The fidelity decays even in the unperturbed case $p=0$ as decribed
above, due to terms that are not included in the Hamiltonian of Eq.
(\ref{eq:hpert-1-1-1}) (see Appendix \ref{sec:Intrinsic-Decoherence-effects.}).
The perturbation $p=0.0017$ represents a limiting case $p\rightarrow0$
from which $f(t)$ no longer improves, indicating that the remaining
decay is originated by the uncontrolled perturbation sources. The
cluster-size $K(t)$ initially grows exponentially as a function of
time and then slows down to a power law behavior whose growth-rate
reduces with increasing perturbation strength. For strong perturbations
$K(t)$ saturates to a value independent of time that decreases with
increasing perturbation strength \citep{alvarez2015localization}.
We call this effect localization of the ``coherence length'' of the
MQC-fidelity that quantifies the ``localization'' of the scrambling
of information shared between the perturbed and ideal dynamics determined
from the OTOC of Eq. (\ref{eq:GOTOC-1}). The fidelity $f(t)$ reaches
an exponential decay regimen with a constant rate when the dynamics
of $K(t)$ is localized (Fig. \ref{fig: f_k_and_chi}(a)). Analogous
results are observed for $\Sigma=\mathcal{H}_{z}$.

\section{Fidelity decay rate scaling with the instantaneous coherence length}

The fidelity decay 
\begin{equation}
f(p,t)=e^{-\chi(p,t)}
\end{equation}
is determined by the instantaneous decoherence rate {[}Fig. \ref{fig: f_k_and_chi}(c){]}
\begin{equation}
\chi'(p,t)=\frac{d\chi}{dt}(p,t).
\end{equation}
For strong perturbations, the decoherence rate $\chi'(t)$ reaches
a plateau \textendash a constant value\textendash{} that depends on
the perturbation strength when the dynamics of $K(t)$ is localized.
However, for weak perturbations when the dynamics of $K(t)$ does
not evidence localization, this plateau is not manifested. Consistently
when localization effects are observed, $\chi'(K)$ evidences an accumulation
of points as shown in Fig. \ref{fig: f_k_and_chi}(d). This demonstrates
that the saturation of $\chi'(t)$ and $K(t)$ occur at the same time.
Moreover, the experimental results show that $\chi'(K)\propto K^{\alpha}$
for long times, indicating that the fidelity decay rate is determined
by a scrambling rate defined by the instantaneous effective cluster-size
of correlated spins $K$.

\begin{figure}
\includegraphics[width=1\columnwidth]{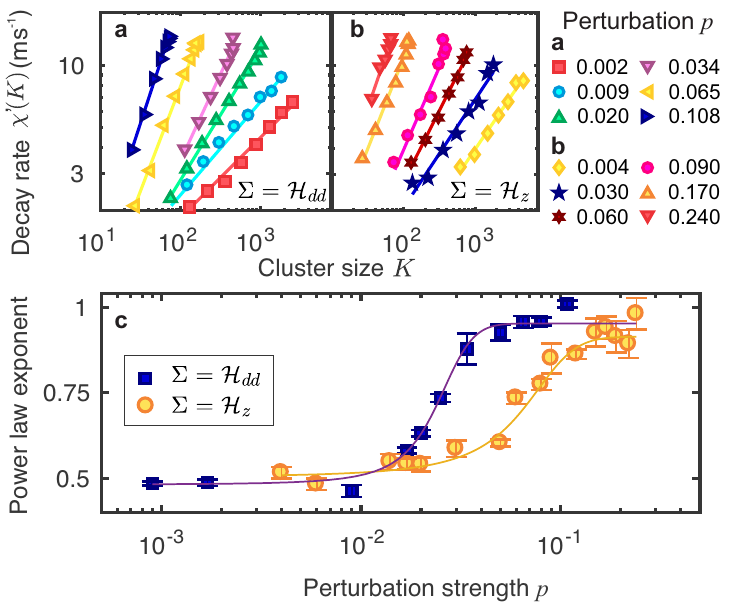}

\caption{Instantaneous decoherence rate $\chi'$ as a function of the effective
cluster-size $K$.\textbf{ }The two perturbation Hamiltonians are
considered (a) $\Sigma=\mathcal{H}_{dd}$ and (b) $\Sigma=H_{z}$.
At long times, the fidelity decay rate is driven by the scrambling
rate $\chi'(K)\sim K^{\alpha}$ given by the instantaneous cluster-size,
with a power-law exponent that depends on the perturbation strength
$p$. (c) The power law exponent $\alpha$ decreases with decreasing
the perturbations strength, showing two plateau values at the weakest
($\alpha_{0}$) and at strongest perturbation ($\alpha_{\infty}$).
To estimate $\alpha_{0}$ and $\alpha_{\infty}$, we fit $\alpha(p)$
with a sigmoid function (solid line). \label{fig:chi_k}}
\end{figure}

Figure \ref{fig:chi_k} shows $\chi'$ as a function of $K$, now
for both perturbation Hamiltonians $\Sigma=\mathcal{H}_{dd}$ and
$\Sigma=\mathcal{H}_{z}$ and different perturbation strengths. The
power law functional form $\chi'(K)\sim K^{\alpha}$ holds for all
the considered cases. The exponents $\alpha$ are shown in Fig. \ref{fig:chi_k}(c).
They give qualitative different limiting values for the localized
(strong perturbation) and delocalized curves (weak perturbation).
For the strongest perturbations, the asymptotic behavior at long
times shows a power law $\chi'(K)\sim K^{\alpha_{\infty}}$, where
$\alpha_{\infty}=0.96\pm0.02$ for the perturbation $\mathcal{H}_{dd}$
and $\alpha_{\infty}=0.91\pm0.03$ for $\mathcal{H}_{z}$, both near
to a linear scaling. However, the exponents drop for the weakest perturbations
as $p\rightarrow0$. In the limit we obtain $\alpha_{0}=(0.48\pm0.03)$
for both $\Sigma$ with the asymptotic behavior $\chi'(K)\sim K^{\alpha_{0}}$.
We expect this exponent to be determined by the uncontrolled perturbation
effects that were not accounted in the experimental quantum simulations
(see Appendix \ref{sec:Intrinsic-Decoherence-effects.}).

\section{Scaling transition on the fidelity decay law: a perturbation threshold}

\begin{figure*}
\includegraphics[width=1\textwidth]{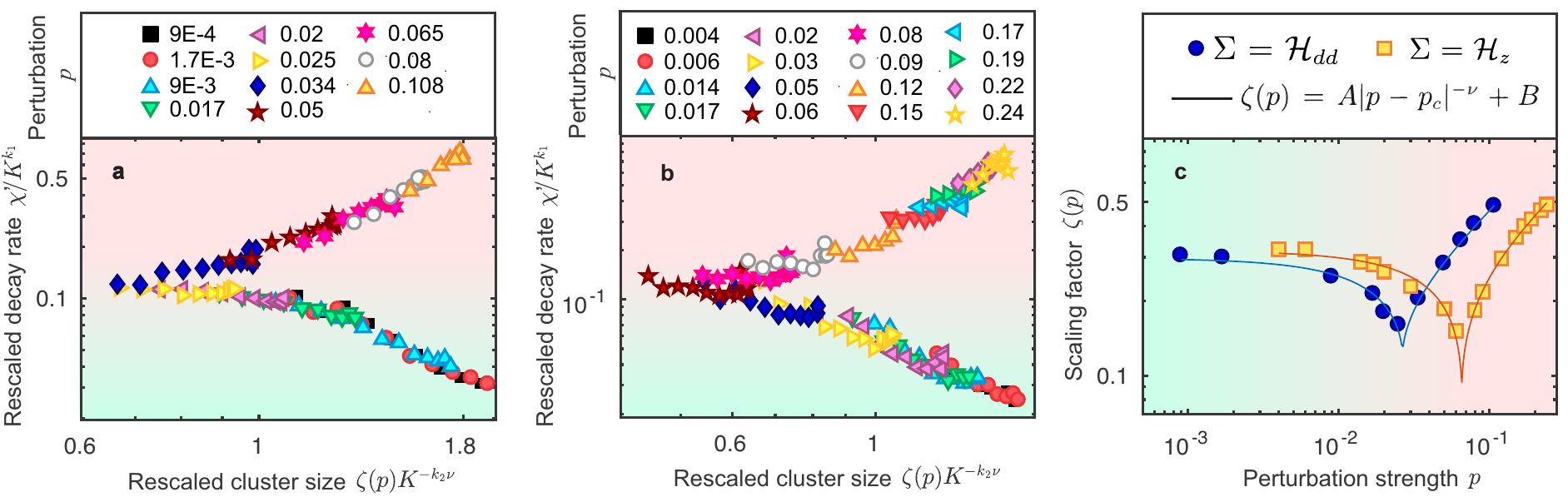}

\caption{\textbf{}Decoherence scaling transition between two dynamical regimes
of the fidelity decay evidenced by the finite-time scaling analysis.
Scalings for both perturbation Hamiltonians (a) $\Sigma=\mathcal{H}_{dd}$
and (b) $\Sigma=\mathcal{H}_{z}$ are shown. (c) The corresponding
scaling factors $\zeta(p)$ and their fittings to the function $\zeta(p)=A|p-p_{c}|^{-\nu}+B$.
In the case of $\Sigma=\mathcal{H}_{dd}$, the critical exponents
are $\nu=(-0.57\pm0.03)$ and $s=(-0.93\pm0.06)$ and the critical
perturbation $p_{c}=(0.026\pm0.001)$. For $\Sigma=\mathcal{H}_{z}$,
the critical perturbation is $p_{c}=(0.065\pm0.01)$, and the critical
exponents are $\nu=(-0.47\pm0.05)$ and $s=(-0.93\pm0.06)$. The curves
of $\zeta(p)$ are normalized to satisfy $\zeta(p_{\infty})=\sqrt{\chi'(K)/K^{\alpha_{\infty}}}$,
where $p_{\infty}=0.108$ for $\Sigma=\mathcal{H}_{dd}$ and $p_{\infty}=0.24$
for $\Sigma=\mathcal{H}_{z}$ are the largest perturbation strength
used in the experiments.\label{fig:Finite-time-scaling-analysis}}
\end{figure*}

To quantitatively analyze the different scaling laws determined by
the exponent $\alpha$, we implement finite-time scaling techniques
typically used to describe localization-delocalization transitions
from finite-time experimental data \citep{chabe2008experimental,lemarie2009observation,alvarez2015localization}.
We consider the evolution time dependence implicit on the cluster-size
$K(t)$. We use the following Ansatz for the scaling behavior at long
times (see Appendix \ref{sec:Finite-time-scaling-procedure.}) 
\begin{equation}
\chi'(p,K)\sim K^{k_{1}}F\left[(p_{c}-p)K^{-k_{2}}\right],\label{eq:scaling_F}
\end{equation}
where $F$ is an arbitrary function. The constants $k_{1}$ and $k_{2}$
are determined to reproduce the asymptotic behavior at weak and strong
perturbations. This assumption leads to the functional regimes $\chi'\sim(p_{c}-p)^{s}K^{\alpha_{0}}$
for $p<p_{c}$ and $\chi'\sim(p-p_{c})^{-2\nu}K^{\alpha_{\infty}}$,
for $p>p_{c}$ at long times. We determine the critical exponents
from the asymptotic experimental data, obtaining $s=(-0.911\pm0.004$),
$\nu=(-0.57\pm0.03)$ for $\Sigma=\mathcal{H}_{dd}$, and $s=(-0.93\pm0.06)$,
$\nu=(-0.47\pm0.05)$ for $\Sigma=\mathcal{H}_{z}$. We then find
the scaling factor $\zeta(p)$ that produces an universal scaling
(see Appendix \ref{sec:Finite-time-scaling-procedure.}).  Rescaled
curves of $\chi'$ as a function of $K$ that collapse into the universal
scaling curve are shown in Fig. \ref{fig:Finite-time-scaling-analysis}(a),(b).
The two branches of the functional behavior, evidence two dynamical
phases for the decoherence effect on the controlled quantum operation
characterized by the scrambling dynamics given by $K(t)$.

The scaling factors $\zeta(p)$ that lead to the universal scalings
for both perturbations are consistent with the single parameter Ansatz
of Eq. (\ref{eq:scaling_F}) that predicts a functional form $\zeta(p)\sim(p-p_{c})^{-\nu}$
(Fig. \ref{fig:Finite-time-scaling-analysis}(c)). The critical perturbation
$p_{c}=(0.026\pm0.001)$ for $\Sigma=\mathcal{H}_{dd}$ is in agreement
with previous experimental values that evidenced a localization-delocalization
transition in the dynamics of the cluster-size $K(t)$ on the same
system\citep{alvarez2015localization}. This coincidence of $p_{c}$
suggests that the critical effects in the dynamics behavior of $K(t)$
and $\chi'(K)$ might be related by a common physical phenomenon.
However, the scaling transition of the exponent $\alpha$ is not determined
by the scaling transition of the dynamic behavior of $K(t)$. The
effect of the localization behavior on $\chi'(K)$ is the accumulation
of points at the end of the curve as shown in Fig. \ref{fig: f_k_and_chi}(d),
but does not determine the power law exponent on the relation $\chi'(K)\sim K^{\alpha}$.

\section{Conclusion}

\textbf{}In summary, we have designed an experiment to quantify the
deviation of a perturbed dynamics from the ideal one based on monitoring
the scrambling of information with MQC and OTOCs. Analyzing many-body
Loschmidt Echoes, we demonstrated that the fidelity decay rate of
the ideal quantum information dynamics is driven by the instantaneous
cluster-size $K(t)$ of correlated spins, which quantifies the information
spreading induced by the control operation. This instantaneous cluster-size
$K(t)$ is an OTOC that gives the common number of correlated spins
shared by the ideal $I_{z}^{0}(t)$ and perturbed $I_{z}(t)$ dynamics.
The fidelity decay shows a transition between two different scaling
laws that depend on the scrambling rate $K^{\alpha}$, whose power
law exponent changes suddenly as a function of the perturbation strength.
By reducing the perturbation strength below a threshold, the exponent
$\alpha$ drops abruptly below 1 and there is no inherent limit to
the number of qubits that can be controlled as expected by the ideal
dynamics. This is encouraging as the dynamical decoherence rate does
not scale linearly with the system size. Although the transition
from one regime to another is smooth due to the finite evolution time
of the experimental data, the finite-time scaling indicates the existence
of the two dynamical regimes. The fact that the controlled dynamics
is more resilient to perturbations if they are below a finite critical
value $p_{c}$, is also promising for allowing reliable quantum control
of large quantum systems. The presented methods provide new avenues
for characterizing the control of many-body systems out-of-equilibrium
with realistic -imperfect- operations for designing novel quantum
technologies.
\begin{acknowledgments}
This work was supported by CNEA, ANPCyT-FONCyT PICT-2017-3447, PICT-2017-3699,
PICT-2018-04333, PIP-CONICET (11220170100486CO), UNCUYO SIIP Tipo
I 2019-C028, Instituto Balseiro. G.A.A. is member of the Research
Career of CONICET. F.D.D. and M.C.R. acknowledges support from CONICET
fellowships.
\end{acknowledgments}

\appendix

\section{Initial state\label{sec:Initial-state.}}

The spin system is described as an ensemble of states with the density
operator. The initial state is a thermal state at room temperature
where $k_{B}T\gg\hbar\omega_{z}$. Therefore, the initial density
matrix is approximated by \citep{slichter2013principles}
\begin{equation}
\rho(0)=\frac{e^{-\frac{\hbar\omega_{z}I_{z}}{k_{B}T}}}{\text{tr\ensuremath{\left\{  e^{-\frac{\hbar\omega_{z}I_{z}}{k_{B}T}}\right\} } }}\approx\left(\mathbb{I}+\frac{\hbar\omega_{z}}{k_{\mathrm{B}}T}I_{z}\right)/\mathrm{Tr}\left\{ \mathbb{I}\right\} .\label{eq:initialstateBoltzmann}
\end{equation}
Notice that as our experimental observable is the spin operator $I_{z}$,
then as $\text{Tr}\left(I_{z}\mathbb{I}\right)=0$, the unity operator
$\mathbb{I}$ in Eq. (\ref{eq:initialstateBoltzmann}) does not contribute
to an observable signal. Then as the NMR signal is $S(t)\propto\text{Tr}\left(I_{z}\rho(t)\right)=\text{Tr}\left(I_{z}U(t)I_{z}U(t)\right)\times\left(\frac{\hbar\omega_{z}}{k_{\mathrm{B}}T}/\mathrm{Tr}\left\{ \mathbb{I}\right\} \right)$,
it gives the time evolution of the operator $I_{z}$ multiplied by
a constant term.

\section{Hamiltonian engineering\label{sec:Hamiltonian-engineering.}}

The effective Hamiltonian of Eq. (\ref{eq:hpert-1-1-1}) is generated
by concatenating short evolution periods $e^{-i\tau_{0}\mathcal{H}_{0}}$
and $e^{-i\tau_{\Sigma}\Sigma}$ of duration $\tau_{0}$ and $\tau_{\Sigma}$
respectively. We get $e^{-i\tau_{0}\mathcal{H}_{0}}e^{-\tau_{\Sigma}\Sigma}=e^{-i\tau_{c}\left[(1-p)\mathcal{H}_{0}+p\Sigma\right]+\mathcal{O}[(\tau_{c}d)^{2}]}$
if the cycle time $\tau_{c}=\tau_{0}+\tau_{\Sigma}\ll d^{-1}$, where
$d\approx13$kHz is the full-width-half-maximum of the resonance
line determined by the homogeneous broadening induced by the dipolar
coupling between the spins. Here $p=\tau_{\Sigma}/\tau_{c}$ is controlled
by adjusting $\tau_{\Sigma}.$ Then based on the Suzuki-Trotter expansion,
the evolution operator $U_{p}(t)$ is achieved by applying repetitively
$N$ cycles $e^{-i\tau_{0}\mathcal{H}_{0}}e^{-\tau_{\Sigma}\Sigma}$
of duration $\tau_{c}$,

\begin{equation}
U_{p}(N\tau_{c})\approx e^{-i\left[(1-p)\mathcal{H}_{0}+p\Sigma\right]N\tau_{c}},
\end{equation}
 where the evolution time $t=N\tau_{c}$.

To engineer the double quantum Hamiltonian $\mathcal{H}_{0}$, we
use the 8-pulse sequence developed in Refs. \citep{Warren1979,baum1985multiple}.
We applied $\pi/2$ RF pulses in the $x$ direction of duration $\tau_{p}=3.24\,\mu\mathrm{s}$,
with delays $\Delta=2\,\mu$s and $\Delta'=2\Delta+\tau_{p}$. The
evolution operator of one cycle is
\begin{multline*}
U_{0}(\tau_{0})=e^{-i\Delta/2\mathcal{H}_{dd}}X^{-1}e^{-i\Delta'\mathcal{H}_{dd}}X^{-1}e^{-i\Delta\mathcal{H}_{dd}}X^{-1}\times\\
\times e^{-i\Delta'\mathcal{H}_{dd}}X^{-1}e^{-i\Delta\mathcal{H}_{dd}}Xe^{-i\Delta'\mathcal{H}_{dd}}Xe^{-i\Delta\mathcal{H}_{dd}}\times\\
\times Xe^{-i\Delta'\mathcal{H}_{dd}}Xe^{-i\Delta/2\mathcal{H}_{dd}},
\end{multline*}
where $X$ is the $\pi/2$-pulse in the $x$ direction. The duration
of the pulse-sequence's cycle in our experiments was $\tau_{0}=62.88\,\mu\mathrm{s}.$
Again if $\tau_{0}d\ll1$, $U(\tau_{0})$ approximates to

\begin{equation}
U_{0}(\tau_{0})\approx e^{-i\tau_{0}\mathcal{H}_{0}}.
\end{equation}
The perturbation $\Sigma=\mathcal{H}_{dd}$ was prepared by a free-evolution
period of duration $\tau_{\Sigma}$ following the cycle of $\mathcal{H}_{0}$
of duration $\tau_{0}$ \citep{alvarez2010nmr,alvarez2015localization}.
The perturbation $\Sigma=\mathcal{H}_{z}=\Delta\omega_{z}I_{z}$ is
produced by phase-shifts of the pulses that generate the $\mathcal{H}_{0}$
Hamiltonian by following the protocol proposed in Ref. \citep{wei2019emergent}.
The $n$-th cycle of the 8-pulse sequence that generates $\mathcal{H}_{0}$
is shifted by an angle $(n-1)\varphi$. Then, the evolution operator
for the $n$-th cycle is

\begin{equation}
U_{n}(\tau_{0})=e^{-iI_{z}(n-1)\varphi}\,e^{-i\mathcal{H}_{0}\tau_{0}}\,e^{iI_{z}(n-1)\varphi},
\end{equation}

\noindent and the concatenation of $N$ cycles is then

\begin{align}
U_{p}(N\,\tau_{0}) & =U_{N}...\,U_{1}\\
 & =e^{-iN\varphi I_{z}}\left[e^{-i\tau_{0}\mathcal{H}_{0}}e^{i\varphi I_{z}}\right]^{N}\nonumber \\
 & =e^{-iN\varphi I_{z}}\left[e^{-i\tau_{0}\mathcal{H}_{0}}e^{i\tau_{\Sigma}dI_{z}}\right]^{N}\\
 & \simeq e^{-iN\varphi I_{z}}e^{-i\tau_{c}N\left[(1-p)\mathcal{H}_{0}+p\Delta\omega_{z}I_{z}\right]},
\end{align}

\noindent where we have defined $\tau_{\Sigma}=\varphi/d$ and $\Delta\omega_{z}=-d$.
As in the case $\Sigma=\mathcal{H}_{dd}$, $p=\tau_{\Sigma}/\tau_{c}$.
The extra phase $e^{-iN\varphi I_{z}}$ is corrected by increasing
the codification phase $\phi$ for determining the MQC spectrum in
an angle $N\varphi$ \citep{wei2019emergent}. The resulting effective
Hamiltonian is then

\begin{equation}
\mathcal{H}(p)\simeq(1-p)\mathcal{H}_{0}+p\mathcal{H}_{z}.
\end{equation}

\section{Fidelity\label{sec:Fidelity.}}

We implement a Loschmidt Echo as a measure of the fidelity between
the ideal density matrix evolving with $U_{0}(t)=e^{-it\mathcal{H}_{0}}$
and the perturbed one evolving with $U_{p}(t)=e^{-it\mathcal{H}(p)}$.
The resulting NMR signal is therefore $S(t)\propto\mathrm{Tr}\left[U_{0}^{\dagger}U_{p}\rho(0)U_{p}^{\dagger}U_{0}\cdot I_{z}\right]\propto\mathrm{Tr}\left[U_{p}I_{z}U_{p}^{\dagger}U_{0}I_{z}U_{0}^{\dagger}\right]=\mathrm{Tr}\left[I_{z}(t)I_{z}^{0}(t)\right]$.
We normalized the experimental data in Fig. \ref{fig: f_k_and_chi}(a)
at $t=0$ to obtain the fidelity $f(t)=S(t)/S(0)=\frac{\mathrm{Tr}\left(I_{z}(t)I_{z}^{0}(t)\right)}{\mathrm{Tr}\left(I_{z}^{2}\right)}$.

\section{Determination of the MQC-spectrum and the MQC-fidelity\label{sec:Determination-of-the}}

The double-quantum Hamiltonian of Eq. (\ref{flip-flip}) flips simultaneously
two spins with the same orientation. Taking into account the selection
rules of the transitions, the $z$-component of the magnetization
$M_{z}$ changes by $M=\Delta M_{z}=\pm2$ to add or subtract $\Delta K=\pm1$
to the number of correlated spins $K$ among which the coherence is
shared \citep{Munowitz1987}. Therefore the change of coherence order
allows to probe the number of correlated spins as a witness of the
quantum information spreading \citep{baum1985multiple}. The spin
density matrix after evolving with the evolution operator $U_{0}(t)$
from the initial state can be decomposed on the coherence orders as

\begin{equation}
\rho(t)=\sum_{M}\sum_{m_{j}-m_{i}=M}\rho_{ij}(t)\left|m_{i}\right\rangle \left\langle m_{j}\right|=\sum_{M}\rho_{M}(t),\label{eq:order_decomp-1}
\end{equation}
where the operator $\rho_{M}(t)=\sum_{m_{j}-m_{i}=M}\rho_{ij}(t)$
contains all the elements of the density operator involving the coherences
of order $M$. Then a rotation $\phi_{z}=e^{-i\phi I_{z}}$ of a phase
$\phi$ around the $z$-axis, changes the density operator to 
\begin{equation}
\rho\left(\phi,t\right)=\phi_{z}\rho(t)\phi_{z}^{-1}=\sum_{M}e^{iM\phi}\rho_{M}(t).
\end{equation}
 The fidelity $f_{\phi}(t)$ with the proper normalization results
then
\begin{align}
f_{\phi}(t) & =\mathrm{Tr}\left[\phi_{z}U_{p}I_{z}U_{p}^{\dagger}\phi_{z}U_{0}\cdot I_{z}U_{0}^{\dagger}\right]/\mathrm{Tr}\left(I_{z}^{2}\right)\\
 & =\mathrm{Tr}\left[I_{z}(\phi,t)I_{z}^{0}(t)\right]/\mathrm{Tr}\left(I_{z}^{2}\right)\\
 & =\sum_{M'}e^{iM\phi}\mathrm{Tr}\left[I_{z,M}(t)I_{z,M}^{0}(t)\right]/\mathrm{Tr}\left(I_{z}^{2}\right)\\
 & =\sum_{M}e^{iM\phi}f_{M},
\end{align}
where $f_{M}(t)=\mathrm{Tr}\left[I_{z,M}(t)I_{z,M}^{0}(t)\right]/\mathrm{Tr}\left(I_{z}^{2}\right)$
is a inner-product that with a proper normalization can be interpreted
as a MQC fidelity. The MQC-fidelity is therefore determined by performing
a Fourier transform on $\phi$ of the echo signal $f_{\phi}(t)$.
Similarly, when $p=0$, $f_{M}(p=0,t)=\mathrm{Tr}\left[I_{z,M}^{0}(t)I_{z,M}^{0}(t)\right]/\mathrm{Tr}\left(I_{z}^{2}\right)$
gives the MQC-spectrum \citep{baum1985multiple}.

\section{OTOCs and the effective cluster-size $K(t)$\label{sec:OTOCs-and-the}}

At $p=0$, the fidelity $f_{\phi}(p=0,t)=\text{Tr}\left[\phi_{z}(t)I_{z}\phi_{z}^{\dagger}(t)I_{z}\right]/\mathrm{Tr}\left(I_{z}^{2}\right)=\left\langle \phi_{z}(t)I_{z}\phi_{z}^{\dagger}(t)I_{z}\right\rangle _{\beta=0}$
is a conventional OTOC, where the expectation value $\left\langle O(t)\right\rangle _{\beta}=\text{Tr}\left\{ O(t)\rho_{\beta}\right\} /\text{Tr}\left\{ O(0)\rho_{\beta}\right\} $
of the operator $O(t)$ is normalized at $t=0$, where $\rho_{\beta}=e^{-\beta\mathcal{H}}/\text{Tr}(e^{-\beta\mathcal{H}})$
is the equilibrium density matrix of the system at the inverse temperature
$\beta$ \citep{maldacena2016bound,Kitaev2018,garttner2018relating,wei2019emergent}.
In our case the OTOC provides information of the system at infinite
temperature with $\beta=0$, i.e. $\rho_{\beta=0}=\mathbb{I}/\text{Tr}(\mathbb{I})$.
The fidelity $f_{\phi}(p=0,t)$ quantifies the degree of non-commutation
of $\phi_{z}(t)$ and $I_{z}$ according to the relation
\begin{equation}
f_{\phi}(p=0,t)=1-\frac{1}{2}\text{Tr}\left([\phi_{z}(t),I_{z}]^{\dagger}[\phi_{z}(t),I_{z}]\right)/\mathrm{Tr}\left(I_{z}^{2}\right).\label{eq:fidelity and otoc for p =00003D 0}
\end{equation}
Performing a Taylor expansion of $f_{\phi}(p=0,t)$ for small $\phi$,
we get the second moment of the MQC-spectrum \citep{khitrin1997growth,garttner2018relating}
\begin{align}
m_{2}^{0}(t) & =\sum_{M}M^{2}f_{M}(p=0,t)\label{eq:secondmoment0}\\
 & =\text{Tr}\left([I_{z}^{0}(t),I_{z}]^{\dagger}[I_{z}^{0}(t),I_{z}]\right)/\mathrm{Tr}\left(I_{z}^{2}\right).
\end{align}
 It is possible to deduce from $m_{2}^{0}$ the number of correlated
spins $K_{0}$ by making assumptions on the MQC spectrum $f_{M}$
\citep{khitrin1997growth}. The most extended model was proposed by
Baum et. al \citep{baum1985multiple,baum1986nmr} that gives a Gaussian
distribution for $f_{M}$ as a function of $M$, where $K_{0}(t)=2m_{2}^{0}(t)$
is determined from the width of the Gaussian distribution. The exact
value of $K_{0}$ will depend on the assumed model for the MQC distribution
\citep{khitrin1997growth}.

When $p\neq0$, $f_{\phi}(t)=\text{Tr}\left(I_{z}(t)\phi_{z}^{\dagger}I_{z}^{0}(t)\phi_{z}\right)/\mathrm{Tr}\left(I_{z}^{2}\right)=\left\langle I_{z}(t)\phi_{z}^{\dagger}I_{z}^{0}(t)\phi_{z}\right\rangle _{\beta=0}$
is a more general OTOC \citep{Kitaev2018} that satisfies

\begin{multline}
f_{\phi}(t)=\\
=\left\{ \text{Tr}\left(I_{z}(t)I_{z}^{0}(t)\right)-\text{\ensuremath{\frac{1}{2}}Tr}\left([I_{z}(t),\phi_{z}]^{\dagger}[I_{z}^{0}(t),\phi_{z}]\right)\right\} /\mathrm{Tr}\left(I_{z}^{2}\right)\\
=f_{\phi=0}(t)-\text{\ensuremath{\frac{1}{2}}Tr}\left([I_{z}(t),\phi_{z}]^{\dagger}[I_{z}^{0}(t),\phi_{z}]\right)/\mathrm{Tr}\left(I_{z}^{2}\right).
\end{multline}
Expanding $f_{\phi}(t)$ in powers of $\phi$, equivalently as was
done for obtaining Eq. (\ref{eq:secondmoment0}), we obtain the second
moment of the MQC distribution $f_{M}(t)$

\begin{align}
m_{2}(t) & =\sum_{M}M^{2}f_{M}(t)\\
 & =\text{Tr}\left([I_{z}(t),I_{z}][I_{z}^{0}(t),I_{z}]^{\dagger}\right)/\mathrm{Tr}\left(I_{z}^{2}\right).
\end{align}
The second moment $m_{2}(t)$ quantifies the overlap between the scrambling
of the ideal evolution $I_{z}^{0}(t)$ and the perturbed evolution
$I_{z}(t)$, determined by the inner-product between the corresponding
commutators $[I_{z}^{0}(t),I_{z}]$ and $[I_{z}(t),I_{z}]$ respectively.
Due to the effect of the perturbation, the total intensity of the
MQC spectrum $\sum_{M}f_{M}(t)=f_{\phi=0}(t)$ decreases as a function
of time, so the second moment $m_{2}$ must be normalized by $f_{\phi=0}(t)$
to determine the width of the MQC distribution. In analogy with the
case $p=0$, the effective number of correlated spins is $K(t)=2m_{2}(t)/f_{\phi=0}(t)$.

\section{Intrinsic Decoherence effects\label{sec:Intrinsic-Decoherence-effects.}}

The ideal form of the effective Hamiltonian $\mathcal{H}_{0}$ of
Eq. (\ref{flip-flip}) is based on an 0-th order approximation using
average Hamiltonian theory \citep{Haeberlen1968}. It can only be
achieved if the dipolar couplings $d_{ij}$ are time independent,
all pulses of the NMR sequences are ideal and the condition $\tau_{c}=\tau_{0}+\tau_{\Sigma}\ll d^{-1}$
is good enough. However, typically these couplings are time dependent
due to thermal fluctuations, and the pulses are not ideal. In addition,
there are non-secular terms neglected in Eq. (\ref{eq:dipolar_interaction}),
and they might also contribute to the quantum dynamics. All these
effects introduce extra terms in the effective Hamiltonian $\mathcal{H}$
of Eq. (\ref{eq:hpert-1-1-1}) and in $\mathcal{H}_{0}$ of Eq. (\ref{flip-flip}).
These extra perturbation terms produce decoherence effects on ms time
scales during the quantum simulations, even for $p=0$. These decoherence
effects reduce the detected signal and the overall fidelity $f(t)$.
Then also the MQC-spectrum is attenuated with an overall global factor.
However, on this study, this decoherence effects do not cause localization
of the information scrambling dynamics on the time scale of our experiments
when $p\rightarrow0$ (see Fig. \ref{fig:chi_k}, black squares).
When $p\ne0$, we quantify the scrambling rate $K$ from the second
moment of Eq. (\ref{eq:GOTOC-1}) generated by $\mathcal{H}_{0}$
after a time-reversed evolution under $\mathcal{-H}_{0}$. This means
that these clusters have survived the decoherence effects. Therefore,
the non-equilibrium many-body dynamics observed by the OTOC of Eq.
(\ref{eq:GOTOC-1}), thus reflects the coherent quantum dynamics generated
by the engineered Hamiltonians. We notice that, the experimentally
observed quantum dynamics occurs over times scales much shorter than
the spin-lattice relaxation time, $T_{1}\approx1$ sec, so we also
neglect the effect of thermalization with the lattice. Therefore,
when the controlled perturbation is set to $p=0$, we consider that
the effective perturbation is not null and we determine the cluster
size of correlated spins using $K(t)=\tfrac{2\sum_{M}M^{2}f_{M}(t)}{\sum_{M}f_{M}(t)}$
as for the $p\ne0$ case.

\section{Finite-time scaling procedure\label{sec:Finite-time-scaling-procedure.}}

To implement the finite-time scaling technique \citep{chabe2008experimental,lemarie2009observation,alvarez2015localization},
we used the asymptotic experimental data for $p\rightarrow0$, that
shows that $\chi'(p\rightarrow0,K)\propto K^{\alpha_{0}}$ for long
times. Then we used the asymptotic experimental data for the largest
perturbation strengths $p_{\infty}=0.108$ for $\Sigma=\mathcal{H}_{dd}$
and $p_{\infty}=0.24$ for $\Sigma=\mathcal{H}_{z}$, that in these
cases $\chi'(p_{\infty},K)\propto K^{\alpha_{\infty}}$ is satisfied
for long times. If there is a transition from these two regimes at
a perturbation $p_{c}$, then close to the transition one expects
a power law dependence on $(p-p_{c})$ for the decoherence rate \citep{chabe2008experimental,lemarie2009observation,alvarez2015localization}.
We then consider the following asymptotic functional dependence at
long times

\begin{equation}
\chi'(p,K)\sim\begin{cases}
(p_{c}-p)^{s}K^{\alpha_{0}} & p<p_{c}\\
(p-p_{c})^{-2\nu}K^{\alpha_{\infty}} & p>p_{c},
\end{cases}\label{eq:chi_pc-2}
\end{equation}
where the time dependence is implicit on $K$.

We use the single-parameter Ansatz for the scaling behavior at long
times in order to find the scaling of the curves of Fig. \ref{fig:chi_k},
consistently with previous experimental findings \citep{alvarez2015localization}

\begin{equation}
\chi'(K,p)\sim K^{k_{1}}F\left[(p_{c}-p)K^{k_{2}}\right].\label{eq:scaling_F-1}
\end{equation}
 Here $F(x)$ is an arbitrary function. Based on the asymptotic behavior
of the experimental data, if $p<p_{c}$, then $\chi'\sim(p_{c}-p)^{s}K^{\alpha_{0}}$,
implying that $F(x)\sim x^{s}$ and

\begin{equation}
k_{1}+sk_{2}=\alpha_{0}.\label{eq:eq1-2}
\end{equation}
Then for $p>p_{c}$, $\chi'\sim(p_{c}-p)^{-2\nu}K^{\alpha_{\infty}}$
implies $F(x)\sim(-x)^{-2\nu}$ and

\begin{equation}
k_{1}-2k_{2}\nu=\alpha_{\infty}.\label{eq:eq2-2}
\end{equation}
 We estimate $p_{c}$ from Fig. \ref{fig:chi_k}(c) and we found that
the experimental data satisfy these asymptotic limits for $p\leq0.009$
and $p\geq0.05$ for $\mathcal{H}_{dd}$, and for $p\leq0.05$ and
$p\geq0.17$ for $\mathcal{H}_{z}$. We obtain $s=(-0.911\pm0.004$),
$\nu=(-0.57\pm0.03)$ for $\Sigma=\mathcal{H}_{dd}$, and $s=(-0.93\pm0.06)$,
$\nu=(-0.47\pm0.05)$ for $\Sigma=\mathcal{H}_{z}$.

The scaling hypothesis is then generalized to 
\begin{equation}
\chi'(K,p)\sim K^{k_{1}}\Phi\left[\zeta(p)K^{-k_{2}\nu}\right]\label{eq:scaling_f}
\end{equation}
 for accounting for the intermediate time regimes, where $\Phi(x)$
and $\zeta(p)$ again are arbitrary functions. This equation is less
restrictive than Eq. (\ref{eq:scaling_F-1}) but includes it. Using
the obtained critical exponents, and the values of $\alpha_{0}$ and
$\alpha_{\infty}$ obtained from the asymptotic limits in Fig. \ref{fig:chi_k}(c),
we get $k_{1}=0.69\pm0.05$ and $k_{2}\nu=-0.13\pm0.02$ for $\Sigma=\mathcal{H}_{dd}$
and $k_{1}=0.69\pm0.07$ and $k_{2}\nu=-0.11\pm0.02$ for $\Sigma=\mathcal{H}_{z}$
from Eqs. (\ref{eq:eq1-2}) and (\ref{eq:eq2-2}). The scaling behavior
is then found by a proper determination of $\zeta(p)$.

To find the scaling factor $\zeta(p)$, we plot the curves of $\frac{\chi'}{K^{k_{1}}}$
as a function of $K^{-k_{2}\nu}$, and shift them by $\zeta(p)$ to
overlap with each other for different values of $p$ in such a way
that they generate a single curve as in Fig. \ref{fig:Finite-time-scaling-analysis}.
A single curve is only obtained if the experimental data is consistent
with the scaling assumptions. To assure the consistency of the scaling
determination, according to Eqs. (\ref{eq:scaling_F}) and (\ref{eq:scaling_f}),
then the scaling factor must satisfy

\begin{equation}
\zeta(p)\sim(p-p_{c})^{-\nu}.
\end{equation}
The curves of $\zeta(p)$ are normalized to satisfy $\zeta(p_{\infty})=\sqrt{\chi'(K)/K^{\alpha_{\infty}}}$,
for the largest perturbation strength used in the experiments $p_{\infty}=0.108$
for $\Sigma=\mathcal{H}_{dd}$ and $p_{\infty}=0.24$ for $\Sigma=\mathcal{H}_{z}$.
We then fit the experimental data with the function $\zeta(p)=A|p-p_{c}|^{-\nu}+B$,
where the parameter $B$ accounts for the finite time experimental
data and external decoherence process that smooth the transition \citep{chabe2008experimental,lemarie2009observation,alvarez2015localization}.
We observed the consistency of the fitted curves and the extracted
critical exponents with the assumed single-parameter Ansatz of Eq.
(\ref{eq:scaling_F}). The critical perturbations from these fittings
are then $p_{c}=(0.026\pm0.006)$ and $(0.065\pm0.01)$ for $\mathcal{H}_{dd}$
and $\mathcal{H}_{z}$ respectively. These values are consistent with
the ones estimated from Fig. \ref{fig:chi_k}(c).

We emphasize that the critical behavior of the scaling exponent $\alpha$
is a new physical phenomenon, which cannot be deduced from the localized-delocalized
transition previously reported in $K(t)$ \citep{alvarez2015localization}.
When the decoherence rate $\chi'$ is parameterized as a function
of time, then localization of $K(t)$ implies localization of $\chi'(t),$
as it is shown in Fig. \ref{fig: f_k_and_chi}(c). Instead, the decoherence
rate parameterized as a function of the systems size $\chi'(K)\propto K^{\alpha}$
provides an scaling exponent $\alpha$ that is independent of the
temporal behavior of $K$. Thus localization of $K(t)$ has no implications
in $\alpha.$

\bibliographystyle{apsrev4-2}
\bibliography{ref}

\begin{thebibliography}{50}%
\makeatletter
\providecommand \@ifxundefined [1]{%
 \@ifx{#1\undefined}
}%
\providecommand \@ifnum [1]{%
 \ifnum #1\expandafter \@firstoftwo
 \else \expandafter \@secondoftwo
 \fi
}%
\providecommand \@ifx [1]{%
 \ifx #1\expandafter \@firstoftwo
 \else \expandafter \@secondoftwo
 \fi
}%
\providecommand \natexlab [1]{#1}%
\providecommand \enquote  [1]{``#1''}%
\providecommand \bibnamefont  [1]{#1}%
\providecommand \bibfnamefont [1]{#1}%
\providecommand \citenamefont [1]{#1}%
\providecommand \href@noop [0]{\@secondoftwo}%
\providecommand \href [0]{\begingroup \@sanitize@url \@href}%
\providecommand \@href[1]{\@@startlink{#1}\@@href}%
\providecommand \@@href[1]{\endgroup#1\@@endlink}%
\providecommand \@sanitize@url [0]{\catcode `\\12\catcode `\$12\catcode
  `\&12\catcode `\#12\catcode `\^12\catcode `\_12\catcode `\%12\relax}%
\providecommand \@@startlink[1]{}%
\providecommand \@@endlink[0]{}%
\providecommand \url  [0]{\begingroup\@sanitize@url \@url }%
\providecommand \@url [1]{\endgroup\@href {#1}{\urlprefix }}%
\providecommand \urlprefix  [0]{URL }%
\providecommand \Eprint [0]{\href }%
\providecommand \doibase [0]{https://doi.org/}%
\providecommand \selectlanguage [0]{\@gobble}%
\providecommand \bibinfo  [0]{\@secondoftwo}%
\providecommand \bibfield  [0]{\@secondoftwo}%
\providecommand \translation [1]{[#1]}%
\providecommand \BibitemOpen [0]{}%
\providecommand \bibitemStop [0]{}%
\providecommand \bibitemNoStop [0]{.\EOS\space}%
\providecommand \EOS [0]{\spacefactor3000\relax}%
\providecommand \BibitemShut  [1]{\csname bibitem#1\endcsname}%
\let\auto@bib@innerbib\@empty
\bibitem [{\citenamefont {Eisert}\ \emph {et~al.}(2015)\citenamefont {Eisert},
  \citenamefont {Friesdorf},\ and\ \citenamefont {Gogolin}}]{Eisert2015}%
  \BibitemOpen
  \bibfield  {author} {\bibinfo {author} {\bibfnamefont {J.}~\bibnamefont
  {Eisert}}, \bibinfo {author} {\bibfnamefont {M.}~\bibnamefont {Friesdorf}},\
  and\ \bibinfo {author} {\bibfnamefont {C.}~\bibnamefont {Gogolin}},\ }\href
  {https://doi.org/10.1038/nphys3215} {\bibfield  {journal} {\bibinfo
  {journal} {Nat. Phys.}\ }\textbf {\bibinfo {volume} {11}},\ \bibinfo {pages}
  {124} (\bibinfo {year} {2015})}\BibitemShut {NoStop}%
\bibitem [{\citenamefont {Abanin}\ \emph {et~al.}(2019)\citenamefont {Abanin},
  \citenamefont {Altman}, \citenamefont {Bloch},\ and\ \citenamefont
  {Serbyn}}]{Abanin2019}%
  \BibitemOpen
  \bibfield  {author} {\bibinfo {author} {\bibfnamefont {D.~A.}\ \bibnamefont
  {Abanin}}, \bibinfo {author} {\bibfnamefont {E.}~\bibnamefont {Altman}},
  \bibinfo {author} {\bibfnamefont {I.}~\bibnamefont {Bloch}},\ and\ \bibinfo
  {author} {\bibfnamefont {M.}~\bibnamefont {Serbyn}},\ }\href
  {https://doi.org/10.1103/RevModPhys.91.021001} {\bibfield  {journal}
  {\bibinfo  {journal} {Rev. Mod. Phys.}\ }\textbf {\bibinfo {volume} {91}},\
  \bibinfo {pages} {021001} (\bibinfo {year} {2019})}\BibitemShut {NoStop}%
\bibitem [{\citenamefont {Martinez}\ \emph {et~al.}(2016)\citenamefont
  {Martinez}, \citenamefont {Muschik}, \citenamefont {Schindler}, \citenamefont
  {Nigg}, \citenamefont {Erhard}, \citenamefont {Heyl}, \citenamefont {Hauke},
  \citenamefont {Dalmonte}, \citenamefont {Monz}, \citenamefont {Zoller},\ and\
  \citenamefont {Blatt}}]{Martinez2016}%
  \BibitemOpen
  \bibfield  {author} {\bibinfo {author} {\bibfnamefont {E.~A.}\ \bibnamefont
  {Martinez}}, \bibinfo {author} {\bibfnamefont {C.~A.}\ \bibnamefont
  {Muschik}}, \bibinfo {author} {\bibfnamefont {P.}~\bibnamefont {Schindler}},
  \bibinfo {author} {\bibfnamefont {D.}~\bibnamefont {Nigg}}, \bibinfo {author}
  {\bibfnamefont {A.}~\bibnamefont {Erhard}}, \bibinfo {author} {\bibfnamefont
  {M.}~\bibnamefont {Heyl}}, \bibinfo {author} {\bibfnamefont {P.}~\bibnamefont
  {Hauke}}, \bibinfo {author} {\bibfnamefont {M.}~\bibnamefont {Dalmonte}},
  \bibinfo {author} {\bibfnamefont {T.}~\bibnamefont {Monz}}, \bibinfo {author}
  {\bibfnamefont {P.}~\bibnamefont {Zoller}},\ and\ \bibinfo {author}
  {\bibfnamefont {R.}~\bibnamefont {Blatt}},\ }\href
  {https://doi.org/10.1038/nature18318} {\bibfield  {journal} {\bibinfo
  {journal} {Nature}\ }\textbf {\bibinfo {volume} {534}},\ \bibinfo {pages}
  {516} (\bibinfo {year} {2016})}\BibitemShut {NoStop}%
\bibitem [{\citenamefont {Swingle}(2018)}]{Swingle2018}%
  \BibitemOpen
  \bibfield  {author} {\bibinfo {author} {\bibfnamefont {B.}~\bibnamefont
  {Swingle}},\ }\href {https://doi.org/10.1038/s41567-018-0295-5} {\bibfield
  {journal} {\bibinfo  {journal} {Nat. Phys.}\ }\textbf {\bibinfo {volume}
  {14}},\ \bibinfo {pages} {988} (\bibinfo {year} {2018})}\BibitemShut
  {NoStop}%
\bibitem [{\citenamefont {Friis}\ \emph {et~al.}(2018)\citenamefont {Friis},
  \citenamefont {Marty}, \citenamefont {Maier}, \citenamefont {Hempel},
  \citenamefont {Holz{\"a}pfel}, \citenamefont {Jurcevic}, \citenamefont
  {Plenio}, \citenamefont {Huber}, \citenamefont {Roos}, \citenamefont
  {Blatt},\ and\ \citenamefont {Lanyon}}]{Friis2018}%
  \BibitemOpen
  \bibfield  {author} {\bibinfo {author} {\bibfnamefont {N.}~\bibnamefont
  {Friis}}, \bibinfo {author} {\bibfnamefont {O.}~\bibnamefont {Marty}},
  \bibinfo {author} {\bibfnamefont {C.}~\bibnamefont {Maier}}, \bibinfo
  {author} {\bibfnamefont {C.}~\bibnamefont {Hempel}}, \bibinfo {author}
  {\bibfnamefont {M.}~\bibnamefont {Holz{\"a}pfel}}, \bibinfo {author}
  {\bibfnamefont {P.}~\bibnamefont {Jurcevic}}, \bibinfo {author}
  {\bibfnamefont {M.~B.}\ \bibnamefont {Plenio}}, \bibinfo {author}
  {\bibfnamefont {M.}~\bibnamefont {Huber}}, \bibinfo {author} {\bibfnamefont
  {C.}~\bibnamefont {Roos}}, \bibinfo {author} {\bibfnamefont {R.}~\bibnamefont
  {Blatt}},\ and\ \bibinfo {author} {\bibfnamefont {B.}~\bibnamefont
  {Lanyon}},\ }\href {https://doi.org/10.1103/PhysRevX.8.021012} {\bibfield
  {journal} {\bibinfo  {journal} {Phys. Rev. X}\ }\textbf {\bibinfo {volume}
  {8}},\ \bibinfo {pages} {021012} (\bibinfo {year} {2018})}\BibitemShut
  {NoStop}%
\bibitem [{\citenamefont {Lewis-Swan}\ \emph {et~al.}(2019)\citenamefont
  {Lewis-Swan}, \citenamefont {Safavi-Naini}, \citenamefont {Kaufman},\ and\
  \citenamefont {Rey}}]{lewis-swan2019dynamics}%
  \BibitemOpen
  \bibfield  {author} {\bibinfo {author} {\bibfnamefont {R.~J.}\ \bibnamefont
  {Lewis-Swan}}, \bibinfo {author} {\bibfnamefont {A.}~\bibnamefont
  {Safavi-Naini}}, \bibinfo {author} {\bibfnamefont {A.~M.}\ \bibnamefont
  {Kaufman}},\ and\ \bibinfo {author} {\bibfnamefont {A.~M.}\ \bibnamefont
  {Rey}},\ }\href {https://doi.org/10.1038/s42254-019-0090-y} {\bibfield
  {journal} {\bibinfo  {journal} {Nat. Rev. Phys.}\ }\textbf {\bibinfo {volume}
  {1}},\ \bibinfo {pages} {627} (\bibinfo {year} {2019})}\BibitemShut {NoStop}%
\bibitem [{\citenamefont {Zhang}\ \emph {et~al.}(2017)\citenamefont {Zhang},
  \citenamefont {Pagano}, \citenamefont {Hess}, \citenamefont {Kyprianidis},
  \citenamefont {Becker}, \citenamefont {Kaplan}, \citenamefont {Gorshkov},
  \citenamefont {Gong},\ and\ \citenamefont {Monroe}}]{Zhang2017}%
  \BibitemOpen
  \bibfield  {author} {\bibinfo {author} {\bibfnamefont {J.}~\bibnamefont
  {Zhang}}, \bibinfo {author} {\bibfnamefont {G.}~\bibnamefont {Pagano}},
  \bibinfo {author} {\bibfnamefont {P.~W.}\ \bibnamefont {Hess}}, \bibinfo
  {author} {\bibfnamefont {A.}~\bibnamefont {Kyprianidis}}, \bibinfo {author}
  {\bibfnamefont {P.}~\bibnamefont {Becker}}, \bibinfo {author} {\bibfnamefont
  {H.}~\bibnamefont {Kaplan}}, \bibinfo {author} {\bibfnamefont {A.~V.}\
  \bibnamefont {Gorshkov}}, \bibinfo {author} {\bibfnamefont {Z.-X.}\
  \bibnamefont {Gong}},\ and\ \bibinfo {author} {\bibfnamefont
  {C.}~\bibnamefont {Monroe}},\ }\href {https://doi.org/10.1038/nature24654}
  {\bibfield  {journal} {\bibinfo  {journal} {Nature}\ }\textbf {\bibinfo
  {volume} {551}},\ \bibinfo {pages} {601} (\bibinfo {year}
  {2017})}\BibitemShut {NoStop}%
\bibitem [{\citenamefont {Bernien}\ \emph {et~al.}(2017)\citenamefont
  {Bernien}, \citenamefont {Schwartz}, \citenamefont {Keesling}, \citenamefont
  {Levine}, \citenamefont {Omran}, \citenamefont {Pichler}, \citenamefont
  {Choi}, \citenamefont {Zibrov}, \citenamefont {Endres}, \citenamefont
  {Greiner}, \citenamefont {Vuleti{\'c}},\ and\ \citenamefont
  {Lukin}}]{Bernien2017}%
  \BibitemOpen
  \bibfield  {author} {\bibinfo {author} {\bibfnamefont {H.}~\bibnamefont
  {Bernien}}, \bibinfo {author} {\bibfnamefont {S.}~\bibnamefont {Schwartz}},
  \bibinfo {author} {\bibfnamefont {A.}~\bibnamefont {Keesling}}, \bibinfo
  {author} {\bibfnamefont {H.}~\bibnamefont {Levine}}, \bibinfo {author}
  {\bibfnamefont {A.}~\bibnamefont {Omran}}, \bibinfo {author} {\bibfnamefont
  {H.}~\bibnamefont {Pichler}}, \bibinfo {author} {\bibfnamefont
  {S.}~\bibnamefont {Choi}}, \bibinfo {author} {\bibfnamefont {A.~S.}\
  \bibnamefont {Zibrov}}, \bibinfo {author} {\bibfnamefont {M.}~\bibnamefont
  {Endres}}, \bibinfo {author} {\bibfnamefont {M.}~\bibnamefont {Greiner}},
  \bibinfo {author} {\bibfnamefont {V.}~\bibnamefont {Vuleti{\'c}}},\ and\
  \bibinfo {author} {\bibfnamefont {M.~D.}\ \bibnamefont {Lukin}},\ }\href
  {https://doi.org/10.1038/nature24622} {\bibfield  {journal} {\bibinfo
  {journal} {Nature}\ }\textbf {\bibinfo {volume} {551}},\ \bibinfo {pages}
  {579} (\bibinfo {year} {2017})}\BibitemShut {NoStop}%
\bibitem [{\citenamefont {Neill}\ \emph {et~al.}(2018)\citenamefont {Neill},
  \citenamefont {Roushan}, \citenamefont {Kechedzhi}, \citenamefont {Boixo},
  \citenamefont {Isakov}, \citenamefont {Smelyanskiy}, \citenamefont {Megrant},
  \citenamefont {Chiaro}, \citenamefont {Dunsworth}, \citenamefont {Arya},
  \citenamefont {Barends}, \citenamefont {Burkett}, \citenamefont {Chen},
  \citenamefont {Chen}, \citenamefont {Fowler}, \citenamefont {Foxen},
  \citenamefont {Giustina}, \citenamefont {Graff}, \citenamefont {Jeffrey},
  \citenamefont {Huang}, \citenamefont {Kelly}, \citenamefont {Klimov},
  \citenamefont {Lucero}, \citenamefont {Mutus}, \citenamefont {Neeley},
  \citenamefont {Quintana}, \citenamefont {Sank}, \citenamefont {Vainsencher},
  \citenamefont {Wenner}, \citenamefont {White}, \citenamefont {Neven},\ and\
  \citenamefont {Martinis}}]{Neill2018}%
  \BibitemOpen
  \bibfield  {author} {\bibinfo {author} {\bibfnamefont {C.}~\bibnamefont
  {Neill}}, \bibinfo {author} {\bibfnamefont {P.}~\bibnamefont {Roushan}},
  \bibinfo {author} {\bibfnamefont {K.}~\bibnamefont {Kechedzhi}}, \bibinfo
  {author} {\bibfnamefont {S.}~\bibnamefont {Boixo}}, \bibinfo {author}
  {\bibfnamefont {S.~V.}\ \bibnamefont {Isakov}}, \bibinfo {author}
  {\bibfnamefont {V.}~\bibnamefont {Smelyanskiy}}, \bibinfo {author}
  {\bibfnamefont {A.}~\bibnamefont {Megrant}}, \bibinfo {author} {\bibfnamefont
  {B.}~\bibnamefont {Chiaro}}, \bibinfo {author} {\bibfnamefont
  {A.}~\bibnamefont {Dunsworth}}, \bibinfo {author} {\bibfnamefont
  {K.}~\bibnamefont {Arya}}, \bibinfo {author} {\bibfnamefont {R.}~\bibnamefont
  {Barends}}, \bibinfo {author} {\bibfnamefont {B.}~\bibnamefont {Burkett}},
  \bibinfo {author} {\bibfnamefont {Y.}~\bibnamefont {Chen}}, \bibinfo {author}
  {\bibfnamefont {Z.}~\bibnamefont {Chen}}, \bibinfo {author} {\bibfnamefont
  {A.}~\bibnamefont {Fowler}}, \bibinfo {author} {\bibfnamefont
  {B.}~\bibnamefont {Foxen}}, \bibinfo {author} {\bibfnamefont
  {M.}~\bibnamefont {Giustina}}, \bibinfo {author} {\bibfnamefont
  {R.}~\bibnamefont {Graff}}, \bibinfo {author} {\bibfnamefont
  {E.}~\bibnamefont {Jeffrey}}, \bibinfo {author} {\bibfnamefont
  {T.}~\bibnamefont {Huang}}, \bibinfo {author} {\bibfnamefont
  {J.}~\bibnamefont {Kelly}}, \bibinfo {author} {\bibfnamefont
  {P.}~\bibnamefont {Klimov}}, \bibinfo {author} {\bibfnamefont
  {E.}~\bibnamefont {Lucero}}, \bibinfo {author} {\bibfnamefont
  {J.}~\bibnamefont {Mutus}}, \bibinfo {author} {\bibfnamefont
  {M.}~\bibnamefont {Neeley}}, \bibinfo {author} {\bibfnamefont
  {C.}~\bibnamefont {Quintana}}, \bibinfo {author} {\bibfnamefont
  {D.}~\bibnamefont {Sank}}, \bibinfo {author} {\bibfnamefont {A.}~\bibnamefont
  {Vainsencher}}, \bibinfo {author} {\bibfnamefont {J.}~\bibnamefont {Wenner}},
  \bibinfo {author} {\bibfnamefont {T.~C.}\ \bibnamefont {White}}, \bibinfo
  {author} {\bibfnamefont {H.}~\bibnamefont {Neven}},\ and\ \bibinfo {author}
  {\bibfnamefont {J.~M.}\ \bibnamefont {Martinis}},\ }\href
  {https://doi.org/10.1126/science.aao4309} {\bibfield  {journal} {\bibinfo
  {journal} {Science}\ }\textbf {\bibinfo {volume} {360}},\ \bibinfo {pages}
  {195} (\bibinfo {year} {2018})}\BibitemShut {NoStop}%
\bibitem [{\citenamefont {Krojanski}\ and\ \citenamefont
  {Suter}(2004)}]{krojanski2004scaling}%
  \BibitemOpen
  \bibfield  {author} {\bibinfo {author} {\bibfnamefont {H.~G.}\ \bibnamefont
  {Krojanski}}\ and\ \bibinfo {author} {\bibfnamefont {D.}~\bibnamefont
  {Suter}},\ }\href {https://doi.org/10.1103/PhysRevLett.93.090501} {\bibfield
  {journal} {\bibinfo  {journal} {Phys. Rev. Lett.}\ }\textbf {\bibinfo
  {volume} {93}},\ \bibinfo {pages} {090501} (\bibinfo {year}
  {2004})}\BibitemShut {NoStop}%
\bibitem [{\citenamefont {{\'{A}}lvarez}\ \emph {et~al.}(2015)\citenamefont
  {{\'{A}}lvarez}, \citenamefont {Suter},\ and\ \citenamefont
  {Kaiser}}]{alvarez2015localization}%
  \BibitemOpen
  \bibfield  {author} {\bibinfo {author} {\bibfnamefont {G.~A.}\ \bibnamefont
  {{\'{A}}lvarez}}, \bibinfo {author} {\bibfnamefont {D.}~\bibnamefont
  {Suter}},\ and\ \bibinfo {author} {\bibfnamefont {R.}~\bibnamefont
  {Kaiser}},\ }\href {https://doi.org/10.1126/science.1261160} {\bibfield
  {journal} {\bibinfo  {journal} {Science}\ }\textbf {\bibinfo {volume}
  {349}},\ \bibinfo {pages} {846} (\bibinfo {year} {2015})}\BibitemShut
  {NoStop}%
\bibitem [{\citenamefont {Suter}\ and\ \citenamefont
  {\'Alvarez}(2016)}]{suter2016colloquium}%
  \BibitemOpen
  \bibfield  {author} {\bibinfo {author} {\bibfnamefont {D.}~\bibnamefont
  {Suter}}\ and\ \bibinfo {author} {\bibfnamefont {G.~A.}\ \bibnamefont
  {\'Alvarez}},\ }\href {https://doi.org/10.1103/RevModPhys.88.041001}
  {\bibfield  {journal} {\bibinfo  {journal} {Rev. Mod. Phys.}\ }\textbf
  {\bibinfo {volume} {88}},\ \bibinfo {pages} {041001} (\bibinfo {year}
  {2016})}\BibitemShut {NoStop}%
\bibitem [{\citenamefont {Wang}\ \emph {et~al.}(2017)\citenamefont {Wang},
  \citenamefont {Um}, \citenamefont {Zhang}, \citenamefont {An}, \citenamefont
  {Lyu}, \citenamefont {Zhang}, \citenamefont {Duan}, \citenamefont {Yum},\
  and\ \citenamefont {Kim}}]{Wang2017}%
  \BibitemOpen
  \bibfield  {author} {\bibinfo {author} {\bibfnamefont {Y.}~\bibnamefont
  {Wang}}, \bibinfo {author} {\bibfnamefont {M.}~\bibnamefont {Um}}, \bibinfo
  {author} {\bibfnamefont {J.}~\bibnamefont {Zhang}}, \bibinfo {author}
  {\bibfnamefont {S.}~\bibnamefont {An}}, \bibinfo {author} {\bibfnamefont
  {M.}~\bibnamefont {Lyu}}, \bibinfo {author} {\bibfnamefont {J.-N.}\
  \bibnamefont {Zhang}}, \bibinfo {author} {\bibfnamefont {L.-M.}\ \bibnamefont
  {Duan}}, \bibinfo {author} {\bibfnamefont {D.}~\bibnamefont {Yum}},\ and\
  \bibinfo {author} {\bibfnamefont {K.}~\bibnamefont {Kim}},\ }\href
  {https://doi.org/10.1038/s41566-017-0007-1} {\bibfield  {journal} {\bibinfo
  {journal} {Nat. Photonics}\ }\textbf {\bibinfo {volume} {11}},\ \bibinfo
  {pages} {646} (\bibinfo {year} {2017})}\BibitemShut {NoStop}%
\bibitem [{\citenamefont {Trotzky}\ \emph {et~al.}(2012)\citenamefont
  {Trotzky}, \citenamefont {Chen}, \citenamefont {Flesch}, \citenamefont
  {McCulloch}, \citenamefont {Schollw{\"o}ck}, \citenamefont {Eisert},\ and\
  \citenamefont {Bloch}}]{Trotzky2012}%
  \BibitemOpen
  \bibfield  {author} {\bibinfo {author} {\bibfnamefont {S.}~\bibnamefont
  {Trotzky}}, \bibinfo {author} {\bibfnamefont {Y.-A.}\ \bibnamefont {Chen}},
  \bibinfo {author} {\bibfnamefont {A.}~\bibnamefont {Flesch}}, \bibinfo
  {author} {\bibfnamefont {I.~P.}\ \bibnamefont {McCulloch}}, \bibinfo {author}
  {\bibfnamefont {U.}~\bibnamefont {Schollw{\"o}ck}}, \bibinfo {author}
  {\bibfnamefont {J.}~\bibnamefont {Eisert}},\ and\ \bibinfo {author}
  {\bibfnamefont {I.}~\bibnamefont {Bloch}},\ }\href
  {https://doi.org/10.1038/nphys2232} {\bibfield  {journal} {\bibinfo
  {journal} {Nat. Phys.}\ }\textbf {\bibinfo {volume} {8}},\ \bibinfo {pages}
  {325} (\bibinfo {year} {2012})}\BibitemShut {NoStop}%
\bibitem [{\citenamefont {Schindler}\ \emph {et~al.}(2013)\citenamefont
  {Schindler}, \citenamefont {M{\"u}ller}, \citenamefont {Nigg}, \citenamefont
  {Barreiro}, \citenamefont {Martinez}, \citenamefont {Hennrich}, \citenamefont
  {Monz}, \citenamefont {Diehl}, \citenamefont {Zoller},\ and\ \citenamefont
  {Blatt}}]{Schindler2013}%
  \BibitemOpen
  \bibfield  {author} {\bibinfo {author} {\bibfnamefont {P.}~\bibnamefont
  {Schindler}}, \bibinfo {author} {\bibfnamefont {M.}~\bibnamefont
  {M{\"u}ller}}, \bibinfo {author} {\bibfnamefont {D.}~\bibnamefont {Nigg}},
  \bibinfo {author} {\bibfnamefont {J.~T.}\ \bibnamefont {Barreiro}}, \bibinfo
  {author} {\bibfnamefont {E.~A.}\ \bibnamefont {Martinez}}, \bibinfo {author}
  {\bibfnamefont {M.}~\bibnamefont {Hennrich}}, \bibinfo {author}
  {\bibfnamefont {T.}~\bibnamefont {Monz}}, \bibinfo {author} {\bibfnamefont
  {S.}~\bibnamefont {Diehl}}, \bibinfo {author} {\bibfnamefont
  {P.}~\bibnamefont {Zoller}},\ and\ \bibinfo {author} {\bibfnamefont
  {R.}~\bibnamefont {Blatt}},\ }\href {https://doi.org/10.1038/nphys2630}
  {\bibfield  {journal} {\bibinfo  {journal} {Nat. Phys.}\ }\textbf {\bibinfo
  {volume} {9}},\ \bibinfo {pages} {361} (\bibinfo {year} {2013})}\BibitemShut
  {NoStop}%
\bibitem [{\citenamefont {Landsman}\ \emph {et~al.}(2019)\citenamefont
  {Landsman}, \citenamefont {Figgatt}, \citenamefont {Schuster}, \citenamefont
  {Linke}, \citenamefont {Yoshida}, \citenamefont {Yao},\ and\ \citenamefont
  {Monroe}}]{Landsman2019}%
  \BibitemOpen
  \bibfield  {author} {\bibinfo {author} {\bibfnamefont {K.~A.}\ \bibnamefont
  {Landsman}}, \bibinfo {author} {\bibfnamefont {C.}~\bibnamefont {Figgatt}},
  \bibinfo {author} {\bibfnamefont {T.}~\bibnamefont {Schuster}}, \bibinfo
  {author} {\bibfnamefont {N.~M.}\ \bibnamefont {Linke}}, \bibinfo {author}
  {\bibfnamefont {B.}~\bibnamefont {Yoshida}}, \bibinfo {author} {\bibfnamefont
  {N.~Y.}\ \bibnamefont {Yao}},\ and\ \bibinfo {author} {\bibfnamefont
  {C.}~\bibnamefont {Monroe}},\ }\href
  {https://doi.org/10.1038/s41586-019-0952-6} {\bibfield  {journal} {\bibinfo
  {journal} {Nature}\ }\textbf {\bibinfo {volume} {567}},\ \bibinfo {pages}
  {61} (\bibinfo {year} {2019})}\BibitemShut {NoStop}%
\bibitem [{\citenamefont {Sar}\ \emph {et~al.}(2012)\citenamefont {Sar},
  \citenamefont {Wang}, \citenamefont {Blok}, \citenamefont {Bernien},
  \citenamefont {Taminiau}, \citenamefont {Toyli}, \citenamefont {Lidar},
  \citenamefont {Awschalom}, \citenamefont {Hanson},\ and\ \citenamefont
  {Dobrovitski}}]{Sar2012}%
  \BibitemOpen
  \bibfield  {author} {\bibinfo {author} {\bibfnamefont {T.~v.~d.}\
  \bibnamefont {Sar}}, \bibinfo {author} {\bibfnamefont {Z.~H.}\ \bibnamefont
  {Wang}}, \bibinfo {author} {\bibfnamefont {M.~S.}\ \bibnamefont {Blok}},
  \bibinfo {author} {\bibfnamefont {H.}~\bibnamefont {Bernien}}, \bibinfo
  {author} {\bibfnamefont {T.~H.}\ \bibnamefont {Taminiau}}, \bibinfo {author}
  {\bibfnamefont {D.~M.}\ \bibnamefont {Toyli}}, \bibinfo {author}
  {\bibfnamefont {D.~A.}\ \bibnamefont {Lidar}}, \bibinfo {author}
  {\bibfnamefont {D.~D.}\ \bibnamefont {Awschalom}}, \bibinfo {author}
  {\bibfnamefont {R.}~\bibnamefont {Hanson}},\ and\ \bibinfo {author}
  {\bibfnamefont {V.~V.}\ \bibnamefont {Dobrovitski}},\ }\href
  {https://doi.org/10.1038/nature10900} {\bibfield  {journal} {\bibinfo
  {journal} {Nature}\ }\textbf {\bibinfo {volume} {484}},\ \bibinfo {pages}
  {82} (\bibinfo {year} {2012})}\BibitemShut {NoStop}%
\bibitem [{\citenamefont {Souza}\ \emph {et~al.}(2012)\citenamefont {Souza},
  \citenamefont {{\'A}lvarez},\ and\ \citenamefont {Suter}}]{Souza2012}%
  \BibitemOpen
  \bibfield  {author} {\bibinfo {author} {\bibfnamefont {A.~M.}\ \bibnamefont
  {Souza}}, \bibinfo {author} {\bibfnamefont {G.~A.}\ \bibnamefont
  {{\'A}lvarez}},\ and\ \bibinfo {author} {\bibfnamefont {D.}~\bibnamefont
  {Suter}},\ }\href {https://doi.org/10.1103/PhysRevA.86.050301} {\bibfield
  {journal} {\bibinfo  {journal} {Phys. Rev. A}\ }\textbf {\bibinfo {volume}
  {86}},\ \bibinfo {pages} {050301(R)} (\bibinfo {year} {2012})}\BibitemShut
  {NoStop}%
\bibitem [{\citenamefont {Taminiau}\ \emph {et~al.}(2014)\citenamefont
  {Taminiau}, \citenamefont {Cramer}, \citenamefont {Sar}, \citenamefont
  {Dobrovitski},\ and\ \citenamefont {Hanson}}]{Taminiau2014}%
  \BibitemOpen
  \bibfield  {author} {\bibinfo {author} {\bibfnamefont {T.~H.}\ \bibnamefont
  {Taminiau}}, \bibinfo {author} {\bibfnamefont {J.}~\bibnamefont {Cramer}},
  \bibinfo {author} {\bibfnamefont {T.~v.~d.}\ \bibnamefont {Sar}}, \bibinfo
  {author} {\bibfnamefont {V.~V.}\ \bibnamefont {Dobrovitski}},\ and\ \bibinfo
  {author} {\bibfnamefont {R.}~\bibnamefont {Hanson}},\ }\href
  {https://doi.org/10.1038/nnano.2014.2} {\bibfield  {journal} {\bibinfo
  {journal} {Nat. Nanotechnol.}\ }\textbf {\bibinfo {volume} {9}},\ \bibinfo
  {pages} {171} (\bibinfo {year} {2014})}\BibitemShut {NoStop}%
\bibitem [{\citenamefont {Zhang}\ and\ \citenamefont
  {Suter}(2015)}]{Zhang2015}%
  \BibitemOpen
  \bibfield  {author} {\bibinfo {author} {\bibfnamefont {J.}~\bibnamefont
  {Zhang}}\ and\ \bibinfo {author} {\bibfnamefont {D.}~\bibnamefont {Suter}},\
  }\href {https://doi.org/10.1103/PhysRevLett.115.110502} {\bibfield  {journal}
  {\bibinfo  {journal} {Phys. Rev. Lett.}\ }\textbf {\bibinfo {volume} {115}},\
  \bibinfo {pages} {110502} (\bibinfo {year} {2015})}\BibitemShut {NoStop}%
\bibitem [{\citenamefont {Schweigler}\ \emph {et~al.}(2017)\citenamefont
  {Schweigler}, \citenamefont {Kasper}, \citenamefont {Erne}, \citenamefont
  {Mazets}, \citenamefont {Rauer}, \citenamefont {Cataldini}, \citenamefont
  {Langen}, \citenamefont {Gasenzer}, \citenamefont {Berges},\ and\
  \citenamefont {Schmiedmayer}}]{Schweigler2017}%
  \BibitemOpen
  \bibfield  {author} {\bibinfo {author} {\bibfnamefont {T.}~\bibnamefont
  {Schweigler}}, \bibinfo {author} {\bibfnamefont {V.}~\bibnamefont {Kasper}},
  \bibinfo {author} {\bibfnamefont {S.}~\bibnamefont {Erne}}, \bibinfo {author}
  {\bibfnamefont {I.}~\bibnamefont {Mazets}}, \bibinfo {author} {\bibfnamefont
  {B.}~\bibnamefont {Rauer}}, \bibinfo {author} {\bibfnamefont
  {F.}~\bibnamefont {Cataldini}}, \bibinfo {author} {\bibfnamefont
  {T.}~\bibnamefont {Langen}}, \bibinfo {author} {\bibfnamefont
  {T.}~\bibnamefont {Gasenzer}}, \bibinfo {author} {\bibfnamefont
  {J.}~\bibnamefont {Berges}},\ and\ \bibinfo {author} {\bibfnamefont
  {J.}~\bibnamefont {Schmiedmayer}},\ }\href
  {https://doi.org/10.1038/nature22310} {\bibfield  {journal} {\bibinfo
  {journal} {Nature}\ }\textbf {\bibinfo {volume} {545}},\ \bibinfo {pages}
  {323} (\bibinfo {year} {2017})}\BibitemShut {NoStop}%
\bibitem [{\citenamefont {Lukin}\ \emph {et~al.}(2019)\citenamefont {Lukin},
  \citenamefont {Rispoli}, \citenamefont {Schittko}, \citenamefont {Tai},
  \citenamefont {Kaufman}, \citenamefont {Choi}, \citenamefont {Khemani},
  \citenamefont {L{\'e}onard},\ and\ \citenamefont {Greiner}}]{Lukin2019}%
  \BibitemOpen
  \bibfield  {author} {\bibinfo {author} {\bibfnamefont {A.}~\bibnamefont
  {Lukin}}, \bibinfo {author} {\bibfnamefont {M.}~\bibnamefont {Rispoli}},
  \bibinfo {author} {\bibfnamefont {R.}~\bibnamefont {Schittko}}, \bibinfo
  {author} {\bibfnamefont {M.~E.}\ \bibnamefont {Tai}}, \bibinfo {author}
  {\bibfnamefont {A.~M.}\ \bibnamefont {Kaufman}}, \bibinfo {author}
  {\bibfnamefont {S.}~\bibnamefont {Choi}}, \bibinfo {author} {\bibfnamefont
  {V.}~\bibnamefont {Khemani}}, \bibinfo {author} {\bibfnamefont
  {J.}~\bibnamefont {L{\'e}onard}},\ and\ \bibinfo {author} {\bibfnamefont
  {M.}~\bibnamefont {Greiner}},\ }\href
  {https://doi.org/10.1126/science.aau0818} {\bibfield  {journal} {\bibinfo
  {journal} {Science}\ }\textbf {\bibinfo {volume} {364}},\ \bibinfo {pages}
  {256} (\bibinfo {year} {2019})}\BibitemShut {NoStop}%
\bibitem [{\citenamefont {Brydges}\ \emph {et~al.}(2019)\citenamefont
  {Brydges}, \citenamefont {Elben}, \citenamefont {Jurcevic}, \citenamefont
  {Vermersch}, \citenamefont {Maier}, \citenamefont {Lanyon}, \citenamefont
  {Zoller}, \citenamefont {Blatt},\ and\ \citenamefont {Roos}}]{Brydges2019}%
  \BibitemOpen
  \bibfield  {author} {\bibinfo {author} {\bibfnamefont {T.}~\bibnamefont
  {Brydges}}, \bibinfo {author} {\bibfnamefont {A.}~\bibnamefont {Elben}},
  \bibinfo {author} {\bibfnamefont {P.}~\bibnamefont {Jurcevic}}, \bibinfo
  {author} {\bibfnamefont {B.}~\bibnamefont {Vermersch}}, \bibinfo {author}
  {\bibfnamefont {C.}~\bibnamefont {Maier}}, \bibinfo {author} {\bibfnamefont
  {B.~P.}\ \bibnamefont {Lanyon}}, \bibinfo {author} {\bibfnamefont
  {P.}~\bibnamefont {Zoller}}, \bibinfo {author} {\bibfnamefont
  {R.}~\bibnamefont {Blatt}},\ and\ \bibinfo {author} {\bibfnamefont {C.~F.}\
  \bibnamefont {Roos}},\ }\href {https://doi.org/10.1126/science.aau4963}
  {\bibfield  {journal} {\bibinfo  {journal} {Science}\ }\textbf {\bibinfo
  {volume} {364}},\ \bibinfo {pages} {260} (\bibinfo {year}
  {2019})}\BibitemShut {NoStop}%
\bibitem [{\citenamefont {Buluta}\ and\ \citenamefont
  {Nori}(2009)}]{buluta2009quantum}%
  \BibitemOpen
  \bibfield  {author} {\bibinfo {author} {\bibfnamefont {I.}~\bibnamefont
  {Buluta}}\ and\ \bibinfo {author} {\bibfnamefont {F.}~\bibnamefont {Nori}},\
  }\href {https://doi.org/10.1126/science.1177838} {\bibfield  {journal}
  {\bibinfo  {journal} {Science}\ }\textbf {\bibinfo {volume} {326}},\ \bibinfo
  {pages} {108} (\bibinfo {year} {2009})}\BibitemShut {NoStop}%
\bibitem [{\citenamefont {Georgescu}\ \emph {et~al.}(2014)\citenamefont
  {Georgescu}, \citenamefont {Ashhab},\ and\ \citenamefont
  {Nori}}]{georgescu2014quantum}%
  \BibitemOpen
  \bibfield  {author} {\bibinfo {author} {\bibfnamefont {I.~M.}\ \bibnamefont
  {Georgescu}}, \bibinfo {author} {\bibfnamefont {S.}~\bibnamefont {Ashhab}},\
  and\ \bibinfo {author} {\bibfnamefont {F.}~\bibnamefont {Nori}},\ }\href
  {https://doi.org/10.1103/RevModPhys.86.153} {\bibfield  {journal} {\bibinfo
  {journal} {Rev. Mod. Phys.}\ }\textbf {\bibinfo {volume} {86}},\ \bibinfo
  {pages} {153} (\bibinfo {year} {2014})}\BibitemShut {NoStop}%
\bibitem [{\citenamefont {Baum}\ \emph {et~al.}(1985)\citenamefont {Baum},
  \citenamefont {Munowitz}, \citenamefont {Garroway},\ and\ \citenamefont
  {Pines}}]{baum1985multiple}%
  \BibitemOpen
  \bibfield  {author} {\bibinfo {author} {\bibfnamefont {J.}~\bibnamefont
  {Baum}}, \bibinfo {author} {\bibfnamefont {M.}~\bibnamefont {Munowitz}},
  \bibinfo {author} {\bibfnamefont {A.~N.}\ \bibnamefont {Garroway}},\ and\
  \bibinfo {author} {\bibfnamefont {A.}~\bibnamefont {Pines}},\ }\href
  {https://doi.org/10.1063/1.449344} {\bibfield  {journal} {\bibinfo  {journal}
  {J. Chem. Phys.}\ }\textbf {\bibinfo {volume} {83}},\ \bibinfo {pages} {2015}
  (\bibinfo {year} {1985})}\BibitemShut {NoStop}%
\bibitem [{\citenamefont {Garttner}\ \emph {et~al.}(2017)\citenamefont
  {Garttner}, \citenamefont {Bohnet}, \citenamefont {Safavi-Naini},
  \citenamefont {Wall}, \citenamefont {Bollinger},\ and\ \citenamefont
  {Rey}}]{garttner2017measuring}%
  \BibitemOpen
  \bibfield  {author} {\bibinfo {author} {\bibfnamefont {M.}~\bibnamefont
  {Garttner}}, \bibinfo {author} {\bibfnamefont {J.~G.}\ \bibnamefont
  {Bohnet}}, \bibinfo {author} {\bibfnamefont {A.}~\bibnamefont
  {Safavi-Naini}}, \bibinfo {author} {\bibfnamefont {M.~L.}\ \bibnamefont
  {Wall}}, \bibinfo {author} {\bibfnamefont {J.~J.}\ \bibnamefont
  {Bollinger}},\ and\ \bibinfo {author} {\bibfnamefont {A.~M.}\ \bibnamefont
  {Rey}},\ }\href {https://doi.org/10.1038/NPHYS4119} {\bibfield  {journal}
  {\bibinfo  {journal} {Nat. Phys.}\ }\textbf {\bibinfo {volume} {13}},\
  \bibinfo {pages} {781} (\bibinfo {year} {2017})}\BibitemShut {NoStop}%
\bibitem [{\citenamefont {G{\"{a}}rttner}\ \emph {et~al.}(2018)\citenamefont
  {G{\"{a}}rttner}, \citenamefont {Hauke},\ and\ \citenamefont
  {Rey}}]{garttner2018relating}%
  \BibitemOpen
  \bibfield  {author} {\bibinfo {author} {\bibfnamefont {M.}~\bibnamefont
  {G{\"{a}}rttner}}, \bibinfo {author} {\bibfnamefont {P.}~\bibnamefont
  {Hauke}},\ and\ \bibinfo {author} {\bibfnamefont {A.~M.}\ \bibnamefont
  {Rey}},\ }\href {https://doi.org/10.1103/PhysRevLett.120.040402} {\bibfield
  {journal} {\bibinfo  {journal} {Phys. Rev. Lett.}\ }\textbf {\bibinfo
  {volume} {120}},\ \bibinfo {pages} {040402} (\bibinfo {year}
  {2018})}\BibitemShut {NoStop}%
\bibitem [{\citenamefont {{\'{A}}lvarez}\ and\ \citenamefont
  {Suter}(2010)}]{alvarez2010nmr}%
  \BibitemOpen
  \bibfield  {author} {\bibinfo {author} {\bibfnamefont {G.~A.}\ \bibnamefont
  {{\'{A}}lvarez}}\ and\ \bibinfo {author} {\bibfnamefont {D.}~\bibnamefont
  {Suter}},\ }\href {https://doi.org/10.1103/physrevlett.104.230403} {\bibfield
   {journal} {\bibinfo  {journal} {Phys. Rev. Lett.}\ }\textbf {\bibinfo
  {volume} {104}},\ \bibinfo {pages} {230403} (\bibinfo {year}
  {2010})}\BibitemShut {NoStop}%
\bibitem [{\citenamefont {Peres}(1984)}]{peres1984stability}%
  \BibitemOpen
  \bibfield  {author} {\bibinfo {author} {\bibfnamefont {A.}~\bibnamefont
  {Peres}},\ }\href {https://doi.org/10.1103/PhysRevA.30.1610} {\bibfield
  {journal} {\bibinfo  {journal} {Phys. Rev. A}\ }\textbf {\bibinfo {volume}
  {30}},\ \bibinfo {pages} {1610} (\bibinfo {year} {1984})}\BibitemShut
  {NoStop}%
\bibitem [{\citenamefont {Pastawski}\ \emph {et~al.}(2000)\citenamefont
  {Pastawski}, \citenamefont {Levstein}, \citenamefont {Usaj}, \citenamefont
  {Raya},\ and\ \citenamefont {Hirschinger}}]{pastawski2000nuclear}%
  \BibitemOpen
  \bibfield  {author} {\bibinfo {author} {\bibfnamefont {H.}~\bibnamefont
  {Pastawski}}, \bibinfo {author} {\bibfnamefont {P.}~\bibnamefont {Levstein}},
  \bibinfo {author} {\bibfnamefont {G.}~\bibnamefont {Usaj}}, \bibinfo {author}
  {\bibfnamefont {J.}~\bibnamefont {Raya}},\ and\ \bibinfo {author}
  {\bibfnamefont {J.}~\bibnamefont {Hirschinger}},\ }\href
  {https://doi.org/https://doi.org/10.1016/S0378-4371(00)00146-1} {\bibfield
  {journal} {\bibinfo  {journal} {Physica A Stat. Mech. Appl.}\ }\textbf
  {\bibinfo {volume} {283}},\ \bibinfo {pages} {166} (\bibinfo {year}
  {2000})}\BibitemShut {NoStop}%
\bibitem [{\citenamefont {Jacquod}\ and\ \citenamefont
  {Petitjean}(2009)}]{jacquod2009decoherence}%
  \BibitemOpen
  \bibfield  {author} {\bibinfo {author} {\bibfnamefont {P.}~\bibnamefont
  {Jacquod}}\ and\ \bibinfo {author} {\bibfnamefont {C.}~\bibnamefont
  {Petitjean}},\ }\href {https://doi.org/10.1080/00018730902831009} {\bibfield
  {journal} {\bibinfo  {journal} {Adv. Phys.}\ }\textbf {\bibinfo {volume}
  {58}},\ \bibinfo {pages} {67} (\bibinfo {year} {2009})}\BibitemShut {NoStop}%
\bibitem [{\citenamefont {Yan}\ \emph {et~al.}(2020)\citenamefont {Yan},
  \citenamefont {Cincio},\ and\ \citenamefont {Zurek}}]{yan2019information}%
  \BibitemOpen
  \bibfield  {author} {\bibinfo {author} {\bibfnamefont {B.}~\bibnamefont
  {Yan}}, \bibinfo {author} {\bibfnamefont {L.}~\bibnamefont {Cincio}},\ and\
  \bibinfo {author} {\bibfnamefont {W.~H.}\ \bibnamefont {Zurek}},\ }\href
  {https://doi.org/10.1103/PhysRevLett.124.160603} {\bibfield  {journal}
  {\bibinfo  {journal} {Phys. Rev. Lett.}\ }\textbf {\bibinfo {volume} {124}},\
  \bibinfo {pages} {160603} (\bibinfo {year} {2020})}\BibitemShut {NoStop}%
\bibitem [{\citenamefont {Maldacena}\ \emph {et~al.}(2016)\citenamefont
  {Maldacena}, \citenamefont {Shenker},\ and\ \citenamefont
  {Stanford}}]{maldacena2016bound}%
  \BibitemOpen
  \bibfield  {author} {\bibinfo {author} {\bibfnamefont {J.}~\bibnamefont
  {Maldacena}}, \bibinfo {author} {\bibfnamefont {S.~H.}\ \bibnamefont
  {Shenker}},\ and\ \bibinfo {author} {\bibfnamefont {D.}~\bibnamefont
  {Stanford}},\ }\href {https://doi.org/10.1007/jhep08(2016)106} {\bibfield
  {journal} {\bibinfo  {journal} {J. High Energy Phys.}\ }\textbf {\bibinfo
  {volume} {2016}},\ \bibinfo {pages} {106}}\BibitemShut {NoStop}%
\bibitem [{\citenamefont {Li}\ \emph {et~al.}(2017)\citenamefont {Li},
  \citenamefont {Fan}, \citenamefont {Wang}, \citenamefont {Ye}, \citenamefont
  {Zeng}, \citenamefont {Zhai}, \citenamefont {Peng},\ and\ \citenamefont
  {Du}}]{li2017measuring}%
  \BibitemOpen
  \bibfield  {author} {\bibinfo {author} {\bibfnamefont {J.}~\bibnamefont
  {Li}}, \bibinfo {author} {\bibfnamefont {R.}~\bibnamefont {Fan}}, \bibinfo
  {author} {\bibfnamefont {H.}~\bibnamefont {Wang}}, \bibinfo {author}
  {\bibfnamefont {B.}~\bibnamefont {Ye}}, \bibinfo {author} {\bibfnamefont
  {B.}~\bibnamefont {Zeng}}, \bibinfo {author} {\bibfnamefont {H.}~\bibnamefont
  {Zhai}}, \bibinfo {author} {\bibfnamefont {X.}~\bibnamefont {Peng}},\ and\
  \bibinfo {author} {\bibfnamefont {J.}~\bibnamefont {Du}},\ }\href
  {https://doi.org/10.1103/PhysRevX.7.031011} {\bibfield  {journal} {\bibinfo
  {journal} {Phys. Rev. X}\ }\textbf {\bibinfo {volume} {7}},\ \bibinfo {pages}
  {031011} (\bibinfo {year} {2017})}\BibitemShut {NoStop}%
\bibitem [{\citenamefont {Niknam}\ \emph {et~al.}(2020)\citenamefont {Niknam},
  \citenamefont {Santos},\ and\ \citenamefont {Cory}}]{niknam2018sensitivity}%
  \BibitemOpen
  \bibfield  {author} {\bibinfo {author} {\bibfnamefont {M.}~\bibnamefont
  {Niknam}}, \bibinfo {author} {\bibfnamefont {L.~F.}\ \bibnamefont {Santos}},\
  and\ \bibinfo {author} {\bibfnamefont {D.~G.}\ \bibnamefont {Cory}},\ }\href
  {https://doi.org/10.1103/PhysRevResearch.2.013200} {\bibfield  {journal}
  {\bibinfo  {journal} {Phys. Rev. Res.}\ }\textbf {\bibinfo {volume} {2}},\
  \bibinfo {pages} {13200} (\bibinfo {year} {2020})}\BibitemShut {NoStop}%
\bibitem [{\citenamefont {S\'anchez}\ \emph {et~al.}(2020)\citenamefont
  {S\'anchez}, \citenamefont {Chattah}, \citenamefont {Wei}, \citenamefont
  {Buljubasich}, \citenamefont {Cappellaro},\ and\ \citenamefont
  {Pastawski}}]{sanchez2020perturbation}%
  \BibitemOpen
  \bibfield  {author} {\bibinfo {author} {\bibfnamefont {C.~M.}\ \bibnamefont
  {S\'anchez}}, \bibinfo {author} {\bibfnamefont {A.~K.}\ \bibnamefont
  {Chattah}}, \bibinfo {author} {\bibfnamefont {K.~X.}\ \bibnamefont {Wei}},
  \bibinfo {author} {\bibfnamefont {L.}~\bibnamefont {Buljubasich}}, \bibinfo
  {author} {\bibfnamefont {P.}~\bibnamefont {Cappellaro}},\ and\ \bibinfo
  {author} {\bibfnamefont {H.~M.}\ \bibnamefont {Pastawski}},\ }\href
  {https://doi.org/10.1103/PhysRevLett.124.030601} {\bibfield  {journal}
  {\bibinfo  {journal} {Phys. Rev. Lett.}\ }\textbf {\bibinfo {volume} {124}},\
  \bibinfo {pages} {030601} (\bibinfo {year} {2020})}\BibitemShut {NoStop}%
\bibitem [{\citenamefont {Wei}\ \emph {et~al.}(2019)\citenamefont {Wei},
  \citenamefont {Peng}, \citenamefont {Shtanko}, \citenamefont {Marvian},
  \citenamefont {Lloyd}, \citenamefont {Ramanathan},\ and\ \citenamefont
  {Cappellaro}}]{wei2019emergent}%
  \BibitemOpen
  \bibfield  {author} {\bibinfo {author} {\bibfnamefont {K.~X.}\ \bibnamefont
  {Wei}}, \bibinfo {author} {\bibfnamefont {P.}~\bibnamefont {Peng}}, \bibinfo
  {author} {\bibfnamefont {O.}~\bibnamefont {Shtanko}}, \bibinfo {author}
  {\bibfnamefont {I.}~\bibnamefont {Marvian}}, \bibinfo {author} {\bibfnamefont
  {S.}~\bibnamefont {Lloyd}}, \bibinfo {author} {\bibfnamefont
  {C.}~\bibnamefont {Ramanathan}},\ and\ \bibinfo {author} {\bibfnamefont
  {P.}~\bibnamefont {Cappellaro}},\ }\href
  {https://doi.org/10.1103/PhysRevLett.123.090605} {\bibfield  {journal}
  {\bibinfo  {journal} {Phys. Rev. Lett.}\ }\textbf {\bibinfo {volume} {123}},\
  \bibinfo {pages} {090605} (\bibinfo {year} {2019})}\BibitemShut {NoStop}%
\bibitem [{\citenamefont {Wei}\ \emph {et~al.}(2018)\citenamefont {Wei},
  \citenamefont {Ramanathan},\ and\ \citenamefont
  {Cappellaro}}]{wei2018exploring}%
  \BibitemOpen
  \bibfield  {author} {\bibinfo {author} {\bibfnamefont {K.~X.}\ \bibnamefont
  {Wei}}, \bibinfo {author} {\bibfnamefont {C.}~\bibnamefont {Ramanathan}},\
  and\ \bibinfo {author} {\bibfnamefont {P.}~\bibnamefont {Cappellaro}},\
  }\href {https://doi.org/10.1103/PhysRevLett.120.070501} {\bibfield  {journal}
  {\bibinfo  {journal} {Phys. Rev. Lett.}\ }\textbf {\bibinfo {volume} {120}},\
  \bibinfo {pages} {070501} (\bibinfo {year} {2018})}\BibitemShut {NoStop}%
\bibitem [{\citenamefont {Fan}\ \emph {et~al.}(2017)\citenamefont {Fan},
  \citenamefont {Zhang}, \citenamefont {Shen},\ and\ \citenamefont
  {Zhai}}]{fan2017out}%
  \BibitemOpen
  \bibfield  {author} {\bibinfo {author} {\bibfnamefont {R.}~\bibnamefont
  {Fan}}, \bibinfo {author} {\bibfnamefont {P.}~\bibnamefont {Zhang}}, \bibinfo
  {author} {\bibfnamefont {H.}~\bibnamefont {Shen}},\ and\ \bibinfo {author}
  {\bibfnamefont {H.}~\bibnamefont {Zhai}},\ }\href
  {https://doi.org/https://doi.org/10.1016/j.scib.2017.04.011} {\bibfield
  {journal} {\bibinfo  {journal} {Sci. Bull}\ }\textbf {\bibinfo {volume}
  {62}},\ \bibinfo {pages} {707} (\bibinfo {year} {2017})}\BibitemShut
  {NoStop}%
\bibitem [{\citenamefont {Garc\'{\i}a-Mata}\ \emph {et~al.}(2018)\citenamefont
  {Garc\'{\i}a-Mata}, \citenamefont {Saraceno}, \citenamefont {Jalabert},
  \citenamefont {Roncaglia},\ and\ \citenamefont
  {Wisniacki}}]{garcia2018chaos}%
  \BibitemOpen
  \bibfield  {author} {\bibinfo {author} {\bibfnamefont {I.}~\bibnamefont
  {Garc\'{\i}a-Mata}}, \bibinfo {author} {\bibfnamefont {M.}~\bibnamefont
  {Saraceno}}, \bibinfo {author} {\bibfnamefont {R.~A.}\ \bibnamefont
  {Jalabert}}, \bibinfo {author} {\bibfnamefont {A.~J.}\ \bibnamefont
  {Roncaglia}},\ and\ \bibinfo {author} {\bibfnamefont {D.~A.}\ \bibnamefont
  {Wisniacki}},\ }\href {https://doi.org/10.1103/PhysRevLett.121.210601}
  {\bibfield  {journal} {\bibinfo  {journal} {Phys. Rev. Lett.}\ }\textbf
  {\bibinfo {volume} {121}},\ \bibinfo {pages} {210601} (\bibinfo {year}
  {2018})}\BibitemShut {NoStop}%
\bibitem [{\citenamefont {Slichter}(1990)}]{slichter2013principles}%
  \BibitemOpen
  \bibfield  {author} {\bibinfo {author} {\bibfnamefont {C.~P.}\ \bibnamefont
  {Slichter}},\ }\href {https://doi.org/10.1007/978-3-662-09441-9} {\emph
  {\bibinfo {title} {{Principles of magnetic resonance}}}}\ (\bibinfo
  {publisher} {Springer-Verlag Berlin Heidelberg},\ \bibinfo {year}
  {1990})\BibitemShut {NoStop}%
\bibitem [{\citenamefont {Munowitz}\ \emph {et~al.}(1987)\citenamefont
  {Munowitz}, \citenamefont {Pines},\ and\ \citenamefont
  {Mehring}}]{Munowitz1987}%
  \BibitemOpen
  \bibfield  {author} {\bibinfo {author} {\bibfnamefont {M.}~\bibnamefont
  {Munowitz}}, \bibinfo {author} {\bibfnamefont {A.}~\bibnamefont {Pines}},\
  and\ \bibinfo {author} {\bibfnamefont {M.}~\bibnamefont {Mehring}},\ }\href
  {https://doi.org/10.1063/1.452028} {\bibfield  {journal} {\bibinfo  {journal}
  {J. Chem. Phys.}\ }\textbf {\bibinfo {volume} {86}},\ \bibinfo {pages} {3172}
  (\bibinfo {year} {1987})}\BibitemShut {NoStop}%
\bibitem [{\citenamefont {Chab\'e}\ \emph {et~al.}(2008)\citenamefont
  {Chab\'e}, \citenamefont {Lemari\'e}, \citenamefont {Gr\'emaud},
  \citenamefont {Delande}, \citenamefont {Szriftgiser},\ and\ \citenamefont
  {Garreau}}]{chabe2008experimental}%
  \BibitemOpen
  \bibfield  {author} {\bibinfo {author} {\bibfnamefont {J.}~\bibnamefont
  {Chab\'e}}, \bibinfo {author} {\bibfnamefont {G.}~\bibnamefont {Lemari\'e}},
  \bibinfo {author} {\bibfnamefont {B.}~\bibnamefont {Gr\'emaud}}, \bibinfo
  {author} {\bibfnamefont {D.}~\bibnamefont {Delande}}, \bibinfo {author}
  {\bibfnamefont {P.}~\bibnamefont {Szriftgiser}},\ and\ \bibinfo {author}
  {\bibfnamefont {J.~C.}\ \bibnamefont {Garreau}},\ }\href
  {https://doi.org/10.1103/PhysRevLett.101.255702} {\bibfield  {journal}
  {\bibinfo  {journal} {Phys. Rev. Lett.}\ }\textbf {\bibinfo {volume} {101}},\
  \bibinfo {pages} {255702} (\bibinfo {year} {2008})}\BibitemShut {NoStop}%
\bibitem [{\citenamefont {Lemari\'e}\ \emph {et~al.}(2009)\citenamefont
  {Lemari\'e}, \citenamefont {Chab\'e}, \citenamefont {Szriftgiser},
  \citenamefont {Garreau}, \citenamefont {Gr\'emaud},\ and\ \citenamefont
  {Delande}}]{lemarie2009observation}%
  \BibitemOpen
  \bibfield  {author} {\bibinfo {author} {\bibfnamefont {G.}~\bibnamefont
  {Lemari\'e}}, \bibinfo {author} {\bibfnamefont {J.}~\bibnamefont {Chab\'e}},
  \bibinfo {author} {\bibfnamefont {P.}~\bibnamefont {Szriftgiser}}, \bibinfo
  {author} {\bibfnamefont {J.~C.}\ \bibnamefont {Garreau}}, \bibinfo {author}
  {\bibfnamefont {B.}~\bibnamefont {Gr\'emaud}},\ and\ \bibinfo {author}
  {\bibfnamefont {D.}~\bibnamefont {Delande}},\ }\href
  {https://doi.org/10.1103/PhysRevA.80.043626} {\bibfield  {journal} {\bibinfo
  {journal} {Phys. Rev. A}\ }\textbf {\bibinfo {volume} {80}},\ \bibinfo
  {pages} {043626} (\bibinfo {year} {2009})}\BibitemShut {NoStop}%
\bibitem [{\citenamefont {Warren}\ \emph {et~al.}(1979)\citenamefont {Warren},
  \citenamefont {Sinton}, \citenamefont {Weitekamp},\ and\ \citenamefont
  {Pines}}]{Warren1979}%
  \BibitemOpen
  \bibfield  {author} {\bibinfo {author} {\bibfnamefont {W.}~\bibnamefont
  {Warren}}, \bibinfo {author} {\bibfnamefont {S.}~\bibnamefont {Sinton}},
  \bibinfo {author} {\bibfnamefont {D.}~\bibnamefont {Weitekamp}},\ and\
  \bibinfo {author} {\bibfnamefont {A.}~\bibnamefont {Pines}},\ }\href
  {https://doi.org/10.1103/PhysRevLett.43.1791} {\bibfield  {journal} {\bibinfo
   {journal} {Phys. Rev. Lett.}\ }\textbf {\bibinfo {volume} {43}},\ \bibinfo
  {pages} {1791} (\bibinfo {year} {1979})}\BibitemShut {NoStop}%
\bibitem [{\citenamefont {Kitaev}\ and\ \citenamefont
  {Suh}(2018)}]{Kitaev2018}%
  \BibitemOpen
  \bibfield  {author} {\bibinfo {author} {\bibfnamefont {A.}~\bibnamefont
  {Kitaev}}\ and\ \bibinfo {author} {\bibfnamefont {S.~J.}\ \bibnamefont
  {Suh}},\ }\href {https://doi.org/10.1007/JHEP05(2018)183} {\bibfield
  {journal} {\bibinfo  {journal} {J. High Energ. Phys.}\ }\textbf {\bibinfo
  {volume} {2018}}\bibinfo  {number} { (5)},\ \bibinfo {pages}
  {183}}\BibitemShut {NoStop}%
\bibitem [{\citenamefont {Khitrin}(1997)}]{khitrin1997growth}%
  \BibitemOpen
\bibfield  {number} {  }\bibfield  {author} {\bibinfo {author} {\bibfnamefont
  {A.}~\bibnamefont {Khitrin}},\ }\href
  {https://doi.org/https://doi.org/10.1016/S0009-2614(97)00661-1} {\bibfield
  {journal} {\bibinfo  {journal} {Chem. Phys. Lett.}\ }\textbf {\bibinfo
  {volume} {274}},\ \bibinfo {pages} {217} (\bibinfo {year}
  {1997})}\BibitemShut {NoStop}%
\bibitem [{\citenamefont {Baum}\ and\ \citenamefont
  {Pines}(1986)}]{baum1986nmr}%
  \BibitemOpen
  \bibfield  {author} {\bibinfo {author} {\bibfnamefont {J.}~\bibnamefont
  {Baum}}\ and\ \bibinfo {author} {\bibfnamefont {A.}~\bibnamefont {Pines}},\
  }\href@noop {} {\bibfield  {journal} {\bibinfo  {journal} {J. Am. Chem.
  Soc.}\ }\textbf {\bibinfo {volume} {108}},\ \bibinfo {pages} {7447} (\bibinfo
  {year} {1986})}\BibitemShut {NoStop}%
\bibitem [{\citenamefont {Haeberlen}\ and\ \citenamefont
  {Waugh}(1968)}]{Haeberlen1968}%
  \BibitemOpen
  \bibfield  {author} {\bibinfo {author} {\bibfnamefont {U.}~\bibnamefont
  {Haeberlen}}\ and\ \bibinfo {author} {\bibfnamefont {J.~S.}\ \bibnamefont
  {Waugh}},\ }\href {https://doi.org/10.1103/PhysRev.175.453} {\bibfield
  {journal} {\bibinfo  {journal} {Phys. Rev.}\ }\textbf {\bibinfo {volume}
  {175}},\ \bibinfo {pages} {453} (\bibinfo {year} {1968})}\BibitemShut
  {NoStop}%
\end{thebibliography}%

\end{document}